\documentclass[12pt]{article}
\pdfoutput=1
\usepackage{latexsym}
\usepackage{epsfig,amssymb,euscript,slashed,stmaryrd}
\usepackage{amsmath}
\usepackage{array,calc,epsfig}
\usepackage[linktocpage]{hyperref}
\usepackage{pstricks}
\usepackage{bbm}
\usepackage{fancybox}

\def\be{\begin{equation}}
\def\ee{\end{equation}}
\def\bseq{\begin{subequations}}
\def\eseq{\end{subequations}}

\def\bea{\begin{eqnarray}}
\def\eea{\end{eqnarray}}
\newcommand\bbone{\ensuremath{\mathbbm{1}}}

\def\bseq{\begin{subequations}}
\def\eseq{\end{subequations}}

\numberwithin{equation}{section} %%
\usepackage[]{graphicx}

\def\d {{\rm d}}
\def\cala         {{\cal A}}

\def\calb         {{\cal B}}
\def\calc         {{\cal C}}
\def\cald         {{\cal D}}

\def\calf         {{\cal F}}
\def\calg         {{\cal G}}

\def\calh         {{\cal H}}
\def\cali         {{\cal I}}
\def\calj         {{\cal J}}
\def\calk         {{\cal K}}
\def\call         {{\cal L}}
\def\calm         {{\cal M}}
\def\caln         {{\cal N}}
\def\calo         {{\cal O}}

\def\calu         {{\cal U}}
\def\calv         {{\cal V}}

\def\del          {\partial}
\def\delbar       {\bar\partial}
\def\ii           {{\rm i}}

\def\tr           {\mathop{\rm Tr}}
\def\Re           {{\rm Re\hskip0.1em}}
\def\Im           {{\rm Im\hskip0.1em}}

\def\sqr#1#2{{\vcenter{\vbox{\hrule height.#2pt
 \hbox{\vrule width.#2pt height#1pt \kern#1pt \vrule width.#2pt}\hrule
 height.#2pt}}}}

\def\d{\text{d}}

\def\slashchar#1{\setbox0=\hbox{$#1$}           % set a box for #1
\dimen0=\wd0                                 % and get its size
\setbox1=\hbox{/} \dimen1=\wd1               % get siste of /
\ifdim\dimen0>\dimen1                        % #1 is bigger
\rlap{\hbox to \dimen0{\hfil/\hfil}}      % so center / in box
#1                                        % and print #1
\else                                        % / is bigger
\rlap{\hbox to \dimen1{\hfil$#1$\hfil}}   % so center #1
/                                         % and print /
\fi}

\begin{document}
\font\cmss=cmss10 \font\cmsss=cmss10 at 7pt

 \begin{flushright}{\scriptsize DFPD-2016/TH/2 
 %\\  \scriptsize  preprint2
 }\end{flushright}
\hfill
\vspace{18pt}
\begin{center}
{\LARGE \textbf{Holographic Effective Field Theories}}
\end{center}
%\begin{center}
% {\LARGE \textbf{ and CFT moduli spaces}}
%\end{center}

%
%\begin{flushright}{\scriptsize preprint1 \\  \scriptsize  preprint2}
%\end{flushright}
%\hfill
%\vspace{18pt}
%\begin{center}
%{\Large \textbf{Notes on warped local models and CFT moduli spaces}}
%\end{center}

\vspace{6pt}
\begin{center}
{\textsl{\rm \large Luca Martucci$\,^a$ and Alberto Zaffaroni$\,^b$}}

\vspace{1cm}
\textit{ $^a$ Dipartimento di Fisica ed Astronomia ``Galileo Galilei",  Universit\`a di Padova\\
\& INFN, Sezione di Padova,
Via Marzolo 8, I-35131 Padova, Italy} \\  \vspace{6pt}
\textit{ $^b$ Dipartimento di Fisica, Universit\`a di Milano-Bicocca, \\
\& INFN, Sezione di Milano-Bicocca, I-20126 Milano, Italy
} \\  \vspace{6pt}
\end{center}

%\vspace{1cm}

\vspace{12pt}

\begin{center}
\textbf{Abstract}

\end{center}

\vspace{4pt} {\small
\noindent 
 We derive the four-dimensional low-energy  effective field theory  governing the moduli space 
of strongly coupled superconformal quiver gauge theories associated with  D3-branes at Calabi-Yau conical singularities in the holographic regime of validity.  
We use the dual supergravity description provided by warped  resolved conical geometries with mobile D3-branes. 
Information on the baryonic directions of the moduli space is also obtained by using
wrapped Euclidean D3-branes.  We illustrate our  general results by discussing in detail their application to the Klebanov-Witten model.

\noindent }

\vspace{1cm}

%\noindent {\em Possible comment ..........................................................................................................................................}

\thispagestyle{empty}

%\vfill
%\vskip 5.mm
%\hrule width 5.cm
%\vskip 2.mm
%{\scriptsize
%\noindent e-mails: {\tt massimo.bianchi@roma2.infn.it, andres.collinucci@physik.uni-muenchen.de, luca.martucci@roma2.infn.it
%}}

\newpage

\setcounter{footnote}{0}

\tableofcontents
%
%\newpage

\section{Introduction}

Since its first explicit incarnation in string theory \cite{Maldacena:1997re}, holography  has been realised in a huge number of possible string/M-theory models, which are dual to various strongly coupled theories,  either conformal or not. The correspondence has been  tested and extended in an impressing number of possible ways. However there are still many potential applications of holography to the study of the {\em dynamics} of strongly coupled systems. 

Consider a strongly coupled theory with a non-trivial moduli space of vacua. If at a generic vacuum the only massless states are given by the moduli, one expects  the low-energy physics to be codified by an appropriate effective field theory  for them. In absence of a sufficient number of (super)symmetries, a  purely field-theoretical  identification of such effective field theory  constitutes a general hard problem. For instance, in four-dimensional $\caln=1$ models, while supersymmetry  significantly helps   the evaluation of the F-terms of the effective theory, there is no general clue on how to face the D-terms directly in field theory. 
Holography provides a natural alternative strategy. If the theory admits a holographic  dual, 
one may use it to identify the effective field theory, which we will refer to as the {\em holographic effective field theory (HEFT)}. The aim of the present paper is to systematically explore this opportunity for a certain broad class of string theory holographic models. 

We will focus on the  four-dimensional $\caln=1$ superconformal field theories (CFTs) which can be engineered by placing $N$ D3-branes at the tip of a six-dimensional cone $C(Y)$ over a Sasaki-Einstein space $Y$. Such  theories are microscopically described by  $\caln=1$ quiver gauge theories that RG-flow to a fixed point at which the theories become superconformal. The prototypical example is provided by the Klebanov-Witten (KW) model \cite{Klebanov:1998hh}, which has been generalised in various ways. All these theories have a rich moduli space  of supersymmetric vacua at which some chiral operators get a non-vanishing vacuum expectation value (vev), the conformal symmetry is spontaneously broken and the dynamics is expected to be describable by an $\caln=1$ effective field theory. Thus, they constitute an ideal laboratory to put the above strategy into practice and, indeed,  we will show how to compute their HEFT.

The holographic realisation of the spontaneously broken phases for our class of models has been discussed in  \cite{Klebanov:1999tb,Klebanov:2007us} in the KW model  and
generalized  in \cite{Martelli:2008cm}.   The ten-dimensional metric is most naturally described as a deformation of AdS$_5\times Y$ in Poincar\'e coordinates and contains an internal non-compact warped Calabi-Yau  space $X$. The warping is sourced by $N$ mobile D3-branes, while $X$ is a resolution of $C(Y)$. In particular, the resolution parameters are naturally associated with the vev of certain baryonic operators and one can choose them so that the supergravity description of the internal space $X$ is justified.

The moduli of these string backgrounds clearly provide the holographic counterpart of the moduli of the dual CFT.  These moduli may be regarded as the moduli of a warped flux compactification of the kind described in \cite{GKP} in which the internal space has been eventually decompactified, so to get an infinite  four-dimensional Planck mass. This viewpoint will help us to identify the Lagrangian of the HEFT by  starting from the effective four-dimensional $\caln=1$ supergravity for flux compactifications   found  in \cite{Martucci:2014ska}, which consistently incorporates the effect of fluxes, warping and mobile D3-branes.  

We will also investigate the explicit connection between the chiral fields entering the HEFT and the vevs of the CFT operators, since the latter should be completely determined by the former. In this regard, the baryonic operators are particularly subtle. Still, we will show that a calculation  along the lines of  \cite{Klebanov:2007us}, see also \cite{Martelli:2008cm}, leads to an explicit  general formula for the baryonic vevs in terms of the HEFT chiral fields.  
% {\tt[more on moduli space?]}

Our general results will be explicitly applied to the KW model. We will  identify its HEFT, explaining in some detail the relation with the dual CFT.  This will be sufficient to  illustrate some key  aspects of the general procedure. On the other hand, other models possess  important properties, as for instance the presence of anomalous baryonic symmetries, which are not shared by the KW model. These would require a further in-depth analysis through the investigation of the HEFT of more general explicit models, which we leave to the future.

The paper is organised as follows. In Section 2 we discuss the structure of the supergravity vacua we are interested in, corresponding to D3-branes moving on a smooth non-compact Calabi-Yau. 
In Section 3 we describe the HEFT, introducing the appropriate chiral moduli and presenting  the associated  K\"alher potential. 
We also provide an alternative description of the moduli space in terms of complex-symplectic coordinates. In Section 4 we compare the HEFT
with the dual CFT expectations. In Section 5 we discuss how to extract baryonic vevs from Euclidean D3-branes, along the lines of  \cite{Klebanov:2007us}. In Section 6
we  illustrate our general results by discussing in detail  the Klebanov-Witten model. Section 7 contains some concluding remarks. 
Finally, a series of Appendices containing technical details end the paper.

%%%%%%%%%%%%%%%%%%%%%%%%%%%%%%%%%%%%%%%%%%%%%%%%%%%%%%%%%%%%%%%%%%%%%%%%%%%%%%%%%%%%%%%%%%%%%%%%%%%%
%%%%%%%%%%%%%%%%%%%%%%%%%%%%%%%%%%%%%%%%%%%%%%%%%%%%%%%%%%%%%%%%%%%%%%%%%%%%%%%%%%%%%%%%%%%%%%%%%%%%

\section{Structure and properties of the string vacua}

In this section we describe the general string backgrounds we focus on in the present paper  and discuss the geometrical properties that will be relevant in the following sections.  

\subsection{Supergravity backgrounds}
\label{sec:sugravacua}

In this paper we focus on non-compact  type IIB backgrounds with Einstein-frame metric
\be\label{10dmetric2}
\ell^{-2}_{\rm s}\d s^2_{10}=e^{2A}\d s^2_{\mathbb{M}^{1,3}}+e^{-2A}\,\d s^2_{X}\,,
\ee
where $\d s^2_{\mathbb{M}^{1,3}}$ is the flat four-dimensional Minkowskian metric and we have factorised a dependence on the string length $\ell_{\rm s}=2\pi\sqrt{\alpha'}$ in order to work in  natural string units.

The internal space $X$ is assumed to be a smooth Calabi-Yau that can be obtained by a crepant resolution of a Calabi-Yau cone $C(Y)$ over a Sasaki-Einstein 5-manifold $Y$. The metric on the singular cone $C(Y)$ can be written as 
\be\label{conmetric}
\d r^2+r^2\d s^2_Y\,.
\ee
 The metric on $X$, $\d s^2_{X}$, behaves asymptotically as (\ref{conmetric}) for $r\rightarrow\infty$.  Being a crepant resolution of $C(Y)$, $X$ has the same complex structure  of $C(Y)$ while its K\"ahler structure is different.
The axio-dilaton 
\be
\tau=C_0+\ii e^{-\phi}
\ee takes a fixed constant value, which we can freely choose so that  $\Im\tau\equiv \frac{1}{g_{\rm s}}\gg1$, in order to guarantee the availability of string perturbative regime. 

The non-trivial warp-factor is due to the presence of $N$ mobile D3-branes. In some internal coordinate system $x^m$ ($m=1,\ldots, 6$) on $X$ they are  located at points $x^m_I$, $I=1,\ldots, N$,  and act as sources of the warp-factor, which must solve the equation
\be\label{warpeq}
\Delta e^{-4A}=*_X\sum_{I} \delta^6_I\,.
\ee 
The general solution of this equation is defined only up to a constant. In this paper we are interested in background having an holographically dual SCFT, which can be regarded as the near-horizon limit of solutions describing $N$ D3-branes sitting at the tip of the cone (\ref{conmetric}). The integration constant is then fixed by requiring that for large $r$ $e^{-4A}$ behaves asymptotically as
\be\label{asymwarp}
e^{-4A}\simeq \frac{R^4}{r^4}+\ldots
\ee
with\footnote{We are using the Einstein-frame metric and dimensionless coordinates. $\alpha'$ corrections are better described in the string frame, which has dimensionful curvature radius $R^4_{\rm st}=\ell^4_{\rm s}g_{\rm s}R^4=\frac{\ell^4_{\rm s}g_{\rm s}N}{4\text{vol}(Y)}$.} 
\be\label{radius}
R^4=\frac{N}{4\text{vol}(Y)}\,.
\ee
The self-dual 5-form $F_5$ has  internal components $\ell_{\rm s}^4*_X\d e^{-4A}$ and satisfies the appropriate quantisation condition 
\be
\int_Y F_5=-\ell^4_{\rm s}\,N\,.
\ee

The general solution of (\ref{warpeq}) with such boundary conditions  can be written as
\be\label{warpsol}
e^{-4A(x)}=\sum^N_{I=1}G(x;x_I)\,,
\ee
where $G(x;x')$ is the Green's function associated with the K\"ahler metric $\d s^2_X$.  Notice that 
 \be
G(x;x')\equiv G(x';x)
 \ee
 and for very large $r$ and finite $r'$, $G(x;x')$ approaches the Green's function for the conical metric (\ref{conmetric}) with $x'=0$:
\be
G^{\rm con}(x;0)= \frac{1}{4\text{vol}(Y)}\,\frac{1}{r^4}\,.
\ee
See \cite{Martelli:2007mk} for a discussion on existence and properties of the Green's function on this class of non-compact Calabi-Yau spaces.

\subsection{Topology, couplings and axionic moduli}
\label{sec:top}

The couplings and the closed string axionic moduli of the above class of backgrounds  can be partly identified by purely topological arguments. 
 The topological properties of $X$, which should be regarded as a space with boundary $\del X\equiv Y$,  are discussed in details in  \cite{Martelli:2008cm}.  Here we review some relevant information.

Every five-dimensional Sasaki-Einstein space $Y$ has the following vanishing Betti numbers
\be
b_1(Y)=b_4(Y)=0\,,
\ee
which follows from the fact that $Y$ has positive Ricci curvature and Myers' theorem. 
On the other hand,  it was proven in \cite{caibar} that $X$ has vanishing Betti numbers
\be\label{bettiX}
b_1(X)=b_5(X)=b_6(X)=0\,.
\ee
In addition, we also assume that $X$ has vanishing 
\be\label{b3assumption}
b_3(X)=0\,.
\ee
Such condition, together with (\ref{bettiX}), imply that no four-dimensional particles or domain-walls can be obtained by wrapping  D1-, D3- or D5-branes on one-, three or five-cycles.
% {\tt [interpretation?]}.
%This can be given  a physical holographic interpretation, in terms of the expected absence of stable particles  and domain walls in the dual field theory, to be discussed later on.  

Flat  shifts of the gauge potentials $B_2$, $C_2$ and $C_4$ give rise to (non-dynamical) parameters and (dynamical) closed string moduli characterising the vacua.  Let us start with  $B_2$, $C_2$. Arbitrary flat shifts of these fields are parametrised by $H^{2}(X;\mathbb{R})$, but  integral large gauge transformations make them periodic, so that they actually take values in a $b_2(X)$-dimensional torus.\footnote{\label{foot:RRK} Large gauge transformations of $B_2$ are given by the elements of $H^2(X;\mathbb{Z})$, so that the corresponding $b_2(X)$-dimensional torus is $H^{2}(X;\mathbb{R})/ H^2(X;\mathbb{Z})$. On the other hand, we avoid writing down the precise periodicities of the R-R fields $C_k$ since they are better specified in the  K-theory framework \cite{Moore:1999gb} and so, generically, they cannot be just identified with the corresponding  integral cohomology groups.}  Since $b_1(Y)=b_3(X)=0$, $ H^1(Y;\mathbb{R})=H^3(X,Y;\mathbb{R})=0$  \footnote{Recall that, for any $n$-dimensional manifold $M$ with boundary $\del M$, $ H_k(M,\del M;\mathbb{Z})$ describe equivalence classes of chains in $M$ which can have a non-trivial boundary on $\del M$ while 
   $  H^k(M,\del M;\mathbb{Z})$ can be represented by compactly supported closed $k$-forms, modulo exact forms $\d\Lambda_{k-1}$, with $\Lambda_{k-1}$ compactly supported.
   The (relative) homology groups are related to the (relative) cohomology groups by  Poncar\'e duality and the universal coefficients theorem, see e.g.\ \cite{hatcher2002algebraic}:  $H_k(M;\mathbb{Z})\simeq H^{n-k}(M,\del M;\mathbb{Z})$, $ H_k(M,\del M;\mathbb{Z})\simeq H^{n-k}(M;\mathbb{Z})$ and, modding out the torsion component, $H_{k}(X,\mathbb{Z})_{\rm free}\simeq H^{k}(X,\mathbb{Z})_{\rm free}$.}
 and we can write  the short exact sequence
\be\label{H2split}
0\longrightarrow H^2(X,Y;\mathbb{R}) \longrightarrow H^2(X;\mathbb{R}) \longrightarrow H^2(Y;\mathbb{R}) \longrightarrow 0\,,
\ee
which shows that $H^2(X;\mathbb{R})$ splits into the sum of a `boundary' component $H^2(Y;\mathbb{R})\simeq H_3(Y;\mathbb{R})$ and a `bulk' component $H^2(X,Y;\mathbb{R})\simeq H_4(X;\mathbb{R})$. Hence there are
\be\label{b2split}
b_2(X)=b_3(Y) + b_4(X)
\ee
possible deformations of the complex combination $C_2-\tau B_2$. The deformations counted by $b_3(Y)$ are non-dynamical and combine with the axio-dilaton $\tau$ to give in total $b_3(Y)+1$ free complex parameters distinguishing these backgrounds. They can be measured by integrating $C_2-\tau B_2$ on two-cycles contained in $Y$ and, as we will discuss later, they correspond to the marginal holomorphic gauge couplings in the dual gauge theory. 
On the other hand,  the deformations of $C_2$ and $B_2$ counted by $b_4(X)$ can be considered as  compactly supported and  they give in total $2b_4(X)$ dynamical real moduli. 

Let us now turn to the moduli associated with $C_4$.
%\footnote{\label{foot:C4} We define $C_4$ as the combination of R-R and $B_2$ gauge-forms which couple to a Euclidean D3-brane  wrapping a four-cycle in $X$ through the integral $\int_{\rm E3} C_4$.} 
A first set of such moduli is parametrised by $H^{4}(X;\mathbb{R})$ (up to periodic identifications due to the large gauge transformations). Since $b_3(X)=b_4(Y)=0$ we can write the short exact sequence 
\be\label{short2}
0\longrightarrow H^3(Y;\mathbb{R}) \longrightarrow H^4(X,Y;\mathbb{R}) \longrightarrow H^4(X;\mathbb{R}) \longrightarrow 0\,,
\ee
which tells us that such  $b_4(X)$ flat deformations of $C_4$ can be in fact uplifted to compactly supported ones.   
On the other hand, a key general result of \cite{Martelli:2008cm} is that, with the specific warping boundary condition    (\ref{asymwarp}), 
there are additional $b_3(Y)$  $C_4$-moduli.  They correspond to exact shifts $\Delta C_4=\d\Lambda_3$ which are compactly supported while $\Lambda_3$ is not. Hence  $\Lambda_3|_Y\neq 0$ and $\d\Lambda_3|_Y=0$, so that $\Lambda_3|_Y$ parametrise  the group $H^3(Y;\mathbb{R})$ appearing in (\ref{short2}). 
From (\ref{short2}), we can then conclude that there is a total of 
\be
\dim H^4(X,Y;\mathbb{R})=\dim H_2(X;\mathbb{R})=b_2(X)=b_3(Y)+b_4(X)\,,
\ee
real   $C_4$ moduli. 
 
In the toric case, the crepant resolutions of the toric singular cone $C(Y)$ can be described in terms of the toric diagram \footnote{A singular Calabi-Yau  toric cone $C(Y)$ is described by a  convex rational cone in $\mathbb{R}^3$ generated by $d$ integral vectors  ${\bf w}^A\in\mathbb{Z}^3$ which lie on a plane in $\mathbb{R}^3$. The toric diagram is the convex polygon with integral vertices that is obtained by projecting  the fan on the plane.} which is a convex polygon in the plane with $d$ integral vertices. 
The smooth crepant resolutions $X$ of $C(Y)$ are in one-to-one correspondence with the complete triangulations of the toric diagram, where again all triangles should have integral vertices. If we call $I$ the number of points with integer coordinates enclosed in the toric diagram, $b_3(Y)$ is given by $d-3$, while $b_4(X)$ is given by $I$.

\subsection{K\"ahler moduli and harmonic forms}
\label{sec:moduli}

Because of the assumption (\ref{b3assumption}), the internal K\"ahler space $X$ has no complex structure moduli. On the other hand, according to the existence theorems of \cite{2009arXiv0906.5191G,2009arXiv0912.3946V}, in any class of $H^2(X;\mathbb{R})$ there exists a Ricci flat  K\"ahler form $J$ which has the appropriate asymptotic conical behaviour. This means that we can expand the K\"ahler cohomological class $[J]$ as follows
\be\label{cohodec}
[J]=v^a[\omega_a]\,,
\ee 
where $[\omega_a]$, $a=1,\ldots, b_2(X)$, is a basis of $H^2(X;\mathbb{Z})$. On the one hand, this implies that $\frac{\del[J]}{\del v^a}=[\omega_a]$. 
An infinitesimal variation  $\delta J$ of the K\"ahler form gives a harmonic (1,1) form \cite{Candelas:1990pi}. Hence, there must exist a set of {\em harmonic} (1,1) forms  $\omega_a$ which are representatives of the integral cohomology classes $[\omega_a]\in H^2(X;\mathbb{Z})$ and are such that\footnote{More precisely, one should fix a complex coordinate system, write $J=J_{i\bar\jmath}\,\d z^i\wedge\d\bar z^{\bar\jmath}$, and then identify $\omega_a=\frac{\del J_{i\bar\jmath}}{\del v^a}\,\d z^i\wedge\d\bar z^{\bar\jmath}$.}  
 \be\label{derJva}
 \frac{\del J}{\del v^a}=\omega_a\,.
 \ee
 The quantisation condition  $[\omega_a]\in H^2(X;\mathbb{Z})$ then reads  $\int_C \omega_a\in \mathbb{Z}$  for any two-cycle $C$.
  In particular, by introducing  a  basis of two-cycles $C^a$, we must have 
   \be\label{omegaquant}
N^a{}_b\equiv \int_{C^a} \omega_b \in \mathbb{Z}\,.
 \ee

 In turn, we can write (\ref{cohodec}) in terms of differential forms as follows
 \be\label{Jdec}
 J=J_0+v^a \omega_a\,,
 \ee
 where $J_0$ is an exact (1,1) form. Viceversa, if one knows a general parametrisation of the K\"ahler form $J$, one can vary it  to generate a basis of $b_2(X)$ harmonic forms and  then select the appropriate K\"ahler moduli $v^a$ by imposing  (\ref{derJva}) for a set of harmonic forms $\omega_a$ satisfying the quantisation condition (\ref{omegaquant}). Notice that the forms $\omega_a$, being harmonic, depend on the K\"ahler moduli $v^a$ (while their homology classes do not) as well as $J_0$. Consistency between (\ref{derJva}) and (\ref{Jdec}) requires that
\be\label{omegaderdef}
 \frac{\del J_0}{\del v^a}=- v^b\frac{\del \omega_b}{\del v^a}\,.
\ee
 
Now, because of (\ref{H2split}) (or, rather, its integral counterpart), we should be able to split $\omega_a$ in two sets $\hat\omega_\alpha$ and $\tilde\omega_\sigma$, with $\alpha=1,\dots,b_4(X)$ and $\sigma=1,\ldots,b_3(Y)$, providing a 
basis of harmonic representative of $H^2(X,Y;\mathbb{Z})$ and  of the non-compactly supported elements of $H^2(X;\mathbb{Z})$, respectively.\footnote{\label{foot:forms}Notice that $\hat\omega_\alpha$ span a cononically defined subspace  $H^2(X,Y;\mathbb{Z})\subset H^2(X;\mathbb{Z})$,  while the non-compactly supported basis $\tilde\omega_\sigma$  canonically span only the quotient space $H^2(X;\mathbb{R})/H^2(X,Y;\mathbb{R})$ and so they can be identified at most up to possible mixed redefinitions $\tilde\omega_\sigma\rightarrow  \tilde\omega_\sigma+n^\alpha_\sigma\hat\omega_\alpha$, with $n^\alpha_\sigma\in \mathbb{Z}$. Such redefinition would imply the mixed redefinition $\hat v^\alpha\rightarrow \hat v^\alpha-n^\alpha_\sigma \tilde v^\sigma$ of the K\"ahler moduli.} 
Indeed, it is known \cite{hausel2004} that $H^2(X,Y;\mathbb{Z})$ admits a representation in terms of $L_2$-normalisable harmonic forms, that is, the $b_4(X)$ harmonic forms $\hat\omega_\alpha$ satisfy the normalisation condition
\be\label{L2norm}
\int_X\hat\omega_\alpha\wedge *\hat\omega_\beta<\infty\,.
\ee  
Actually, one can identify the asymptotic behaviour \cite{Martelli:2008cm}
\be\label{asymphatomega}
\|\hat\omega_\alpha\|^2\sim \frac{1}{r^{8+\mu}}
\ee
in the limit $r\rightarrow \infty$, where $\|\hat\omega_\alpha\|^2\equiv \hat\omega_\alpha\lrcorner \hat\omega_\alpha$ and $\mu>0$. Clearly (\ref{asymphatomega}) is compatible with (\ref{L2norm}). 

On the other hand, the  $b_3(Y)$  harmonic forms $\tilde\omega_\sigma$ {\em are not} $L_2$-normalisable. However, by using the fact that $\tilde\omega_\sigma$ asymptotically define a non-trivial element of $H^2(Y;\mathbb{Z})$, one can argue that \cite{Martelli:2008cm}
\be\label{asymphatomega2}
\|\tilde\omega_\sigma\|^2\sim \frac{1}{r^4}\,.
\ee
 This implies that the forms $\tilde\omega_\rho$ are normalisable with respect to 
 the {\em warped} inner product
\be\label{wL2norm}
\int_Xe^{-4A}\tilde\omega_\rho\wedge*\tilde\omega_\sigma<\infty\,.
\ee
We then say that $\tilde\omega_\rho$ are $L_2^{\rm w}$-normalisable. Notice that (\ref{wL2norm})
 is possible  only because of the specific asymptotic behaviour (\ref{asymwarp}) of warping.  
 With an additional constant contribution to $e^{-4A}$, as it would happen in local models of flux compactifications (without taking the near-horizon limit), (\ref{wL2norm}) would not hold anymore.

 An important observation is that all harmonic 2-forms $\omega_a=(\hat\omega_\alpha,\tilde\omega_\sigma)$ are {\em primitive}. Indeed, we can decompose $\omega_a$ in primitive and non-primitive part, $\omega_a=\omega_a^{\rm P}+\alpha_aJ$, so that $\|\omega_a\|^2=\|\omega_a^{\rm P}\|^2+3(\alpha_a)^2$. Consistency with (\ref{asymphatomega}) and (\ref{asymphatomega2}) requires  that $(\alpha_a)^2$ decreases at least as $r^{-4}$. On the other hand $\alpha_a=\frac13 J\lrcorner\omega_a$ is a regular harmonic function, since the contraction with the K\"ahler form $J$ commutes with the Laplacian.  Hence $\alpha_a$ necessarily vanishes and $\omega_a$ is primitive. 
 
 Notice that, of course, the forms $\hat\omega_\alpha$ are $L_2^{\rm w}$-normalisable too, which is consistent with the fact that the forms $\tilde\omega_\sigma$ are defined up to linear combinations of $\hat\omega_\alpha$ (see footnote \ref{foot:forms}). In particular, this implies that  the matrix
\be\label{defcalg}
\calg_{ab}=\int_X e^{-4A}\omega_a\wedge *\omega_b\equiv -\int_X e^{-4A}J\wedge \omega_a\wedge \omega_b
\ee
is well defined and non-degenerate and can be regarded as a positive definite metric on the $b_2(X)$-dimensional  space spanned by the complete set of harmonic forms $\omega_a$. 

 In \cite{caibar} it is shown that $H^2(X;\mathbb{Z})$ is isomorphic to the Picard group of holomorphic line bundles. This implies that the harmonic forms $\omega_a$ can be chosen to be Poincar\'e dual to a basis of divisors $D_a=(\hat D_\alpha,\tilde D_\sigma)$,   which explicitly realise the Poincar\'e duality $H^2(X;\mathbb{Z})\simeq H_{4}(X,Y;\mathbb{Z})$.  In particular, the forms $\hat\omega_\alpha$ are dual to a basis of compact divisors $\hat D_\alpha$, while $\tilde\omega_\sigma$ are dual to non-compact divisors $\tilde D_\sigma$ whose boundary $\del \tilde D_\sigma\subset Y$ define non-trivial non-torsional classes in $H_3(Y;\mathbb{Z})$. Furthermore, the matrix (\ref{omegaquant}) can be represented as intersection matrix $N^a{}_b=C^a\cdot D_b$.

Since the $(1,1)$ form $J_0$ appearing in (\ref{Jdec}) is exact, we can  write it as\footnote{Indeed, we can globally write $J_0=\del\theta^{0,1}+\delbar\bar\theta^{0,1}$ with $\delbar\theta^{0,1}=0$. On the other hand, by Lemma 5.5 of \cite{2009arXiv0906.5191G} we can write $\theta^{0,1}=\delbar f$ for some globally defined function $f$ so that $J_0=\del\delbar f+\delbar\del \bar f=2\ii  \del\delbar \Im f$. We can then set $2\Im f\equiv k_0$ and obtain (\ref{defk0}).} 
\be\label{defk0}
J_0 =\ii\del\delbar k_0\,,
\ee
where  $k_0$ is a globally defined real function.  Notice that $k_0$ depends not only on the coordinates but also on the K\"ahler moduli $v^a$ and then we will sometime more explicitly write $k_0(z,\bar z;v)$.
As we will see, this function plays a crucial role in the description of the low-energy effective theory describing these vacua.

Analogously, we can introduce the potentials $\kappa_a(z,\bar z;v)$ such that
\be
\omega_a=\ii\del\delbar\kappa_a\,.
\ee
Since $\omega_a$ define non-trivial classes in $H^2(X;\mathbb{Z})$, $\kappa_a(z,\bar z;v)$ are  only locally defined. In fact, we can regard $e^{-2\pi\kappa_a}$ as a metric on the line bundle $\calo(D_a)$. More explicitly,  if $\kappa_a(z,\bar z;v)$ has transition functions
\be\label{genkappatr}
\kappa_a(z,\bar z;v)\quad\longrightarrow\quad \kappa_a(z,\bar z;v)+\chi_a(z)+\bar\chi_a(\bar z)\,,
\ee
then a section of the corresponding line bundle $\calo(D_a)$ must transform as
\be\label{genzetatr}
\zeta_a(z)\quad\longrightarrow\quad e^{2\pi\chi_a(z)}\zeta_a(z)\,.
\ee

Notice that $k_0(z,\bar z;v)$, as well as each potential $\kappa_a(z,\bar z;v)$, is defined up to  a $v$-dependent function which does not depend on the coordinates. 
We partially fix such ambiguity by requiring that
\be\label{defk0cond}
\frac{\del k_0}{\del v^a}=-v^b\frac{\del \kappa_b}{\del v^a}\,,
\ee
which is indeed compatible with  (\ref{omegaderdef}). Hence, the asymptotic behaviour of $\frac{\del k_0}{\del v^a}$ is dictated by the asymptotic behaviour of the globally defined functions $\frac{\del \kappa_b}{\del v^a}$, which we fix  as follows.  By adapting to the present context an  
an argument given in \cite{Martucci:2014ska}, we first observe that the primitivity of $\omega_a$ requires, by consistency, that $\frac{\del (J\lrcorner \omega_a)}{\del v^b}=0$. Now, from $\frac{\del J_{i\bar\jmath}}{\del v^a}=(\omega_a)_{i\bar\jmath}$ and $J^{i\bar k}J_{j\bar k}=\delta^i_k$ one can deduce that $\frac{\del J^{i\bar\jmath}}{\del v^a}=-(\omega_a)^{i\bar\jmath}$ and then $\frac{\del (J\lrcorner \omega_a)}{\del v^b}=-\omega_a\lrcorner\omega_b+J\lrcorner \frac{\del\omega_a}{\del v^b}$. On the other hand $J\lrcorner \frac{\del\omega_a}{\del v^b}=J\lrcorner (\ii\del\delbar)\frac{\del\kappa_a}{\del v^b}\equiv -\frac12\Delta  \frac{\del\kappa_a}{\del v^b}$, where $\Delta\equiv -2\ii J\lrcorner \del\delbar$ is the Laplacian associated with the Calabi-Yau metric on $X$, so that we see that the above consistency condition can be written in the form 
\be\label{kappaeq}
\Delta \frac{\del\kappa_a}{\del v^b}=-2\omega_a\lrcorner\omega_b\,.
\ee
This can be immediately integrated by using the Green's function introduced in section \ref{sec:sugravacua}, providing a particular solution of (\ref{kappaeq}) 
%which we impose for $\kappa_a$:
\be\label{intdelkappa}
\frac{\del\kappa_a(x;v)}{\del v^b}=2\int_{X,x'} G(x;x')(J\wedge \omega_a\wedge \omega_b)(x')\,.
\ee
Since $G(x;x')\sim \frac{1}{r'{}^4}$  and $J\wedge \omega_a\wedge \omega_b\equiv -\omega_a\lrcorner\omega_b \d\text{vol}_X$ diverges slower than $r'\d r'\wedge \d\text{vol}_{Y}$ for $r\rightarrow\infty$,  
the integral on the r.h.s.\ of (\ref{intdelkappa}) is indeed well defined. 

Since $\omega_a$ has specific boundary conditions (\ref{asymphatomega}) and (\ref{asymphatomega2}), we see that (\ref{intdelkappa}) implies that $\frac{\del\kappa_a}{\del v^b}$   obey the boundary conditions   
\be\label{asymptkappa}
\frac{\del \kappa_a}{\del v^b}\sim O(r^{-k})\quad~~~\text{with}\ k\geq 2\, .
\ee 
These boundary conditions as well as (\ref{defk0cond})  almost completely fix the possible ambiguity in  $k_0(z,\bar z;v)$ and $\kappa_a(z,\bar z;v)$, so that each of these functions is now uniquely defined up to a possible additive constant.

%%%%%%%%%%%%%%%%%%%%%%%%%%%%%%%%%%%%%%%%%%%%%%%%%%%%%%%%%%%%%%%%%%%%%%%%%%%%%%%%%%%%%%%%%%%%%%%%%%%%
%%%%%%%%%%%%%%%%%%%%%%%%%%%%%%%%%%%%%%%%%%%%%%%%%%%%%%%%%%%%%%%%%%%%%%%%%%%%%%%%%%%%%%%%%%%%%%%%%%%%

%%%%%%%%%%%%%%%%%%%%%%%%%%%%%%%%%%%%%%%%%%%%%%%%%%%%%%%%%%%%%%%%%%%%%%%%%%%%%%%%%%%%%%%%%%%%

\section{The holographic effective field theory}
\label{sec:EFT}

We now turn to the supersymmetric holographic effective field theory (HEFT) describing the low-energy dynamics of the supergravity vacua. Our strategy is to derive it by considering an appropriate rigid limit of the warped supergravities derived in \cite{Martucci:2014ska}. We now explain the logic of this approach, relegating  to appendix \ref{app:rigidlimit}  a more detailed description of the rigid limit, which may be applied to more general non-compact warped F-theory backgrounds.

We start by observing  that the class of holographic backgrounds reviewed in the previous section can be considered as particular subcases of the general class of warped F-theory vacua described  in \cite{GKP}. If the internal space were compact, the four-dimensional low-energy dynamics of the moduli  would be described by an appropriate $\caln=1$ supergravity. In particular, the four-dimensional Planck mass $M_{\rm P}$ would be proportional to the square root of the volume of the internal space, see appendix \ref{app:rigidlimit} for more details. Hence, one may consider our holographic backgrounds as particular rigid limits  of this class of compactifications, in which $M_{\rm P}\rightarrow\infty$ and the internal space decompactifies. In such rigid limit some moduli and their superpartners survive as dynamical fields, i.e.\ their  kinetic terms in the four-dimensional effective theory remain finite. On the other hand, other moduli, as well as the graviton  and their superpartners, acquire an infinite four-dimensional kinetic term, hence ``freezing out''  from the low-energy four-dimensional dynamics. Such decoupled moduli then become non-dynamical parameters in the resulting rigid effective field theory.  

Now, a description of the $\caln=1$ effective supergravity of the warped F-theory vacua of \cite{GKP}, which consistently incorporates the non-trivial  warping and hence the backreaction of fluxes and D3-branes, has been recently provided in \cite{Martucci:2014ska}. Crucially, the relevant quantities appearing in the action can be described in purely geometrical terms. Hence, as discussed in appendix \ref{app:rigidlimit}, one can implement the rigid $M_{\rm P}\rightarrow\infty$ limit at a purely geometrical level, as a decompactification limit, obtaining  geometrical formulas for the resulting rigid four-dimensionl effective theory, which in our context represents the HEFT. As we will review below, the relevant kinetic terms can be expressed in terms of the integrals (\ref{defcalg}) and their unwarped counterpart.  A background modulus  must be then considered a dynamical field of the HEFT if the integral defining the corresponding kinetic term is finite. Otherwise  it is dynamically frozen and parametrises a marginal deformation of the model. 

 In this section we will summarise the main results of rigid limit described in appendix \ref{app:rigidlimit}, showing how the resulting HEFT can be written in a manifestly supersymmetric way. In particular we will describe in detail the appropriate chiral parametrisation of the dynamical moduli and we will identify the K\"ahler potential which defines the HEFT.

\subsection{Chiral moduli}
\label{sec:chiralmoduli}

Let us first organise the spectrum of the moduli in chiral fields. There are $3N$ chiral fields $z^i_I$, $I=1,\ldots,N$, describing the position of the $N$
D3-branes on $X$ in some complex coordinate system $z^i$. In addition, there are the closed string moduli described in the previous section.  All the moduli can be organised in the chiral fields  summarised in the following table
\begin{table}[h!]
\begin{center}
  \begin{tabular}{ | c | c | c | }
    \hline
    chiral fields & moduli  & indices \\ \hline\hline
       $ z^i_I$ & D3's position   &$ i=1,2,3$, $I=1,\ldots,N$ \\
    \hline
       $\beta^\alpha$ & $B_2$, $C_2$   &$\alpha=1,\ldots,b_4(X)$ \\ \hline
     & & $a=1,\ldots,b_2(X)$ \\
    $\rho_a=(\hat\rho_\alpha,\tilde\rho_\sigma)$ & $J$, $C_4$ & $\alpha=1,\ldots,b_4(X)$ \\ 
      & & $\sigma=1,\ldots,b_3(Y)$ \\ \hline
 
  \end{tabular}
  \caption{Chiral fields parametrising open and closed string moduli.}
  \label{table:chiral}
  \end{center} 
\end{table}

The chiral fields $\beta_\alpha$ are obtained by expanding $B_2$ and $C_2$ in the basis of $b_2(X)=b_4(X)+b_2(Y)$ harmonic two-forms $\omega_a=(\hat\omega_\alpha,\tilde\omega_\sigma)$:
\be\label{C2B2exp}
 C_2-\tau B_2=\ell^2_{\rm s}\left(\beta^\alpha\hat\omega_\alpha+\lambda^\sigma \tilde\omega_\sigma\right)\,.
\ee
Here $\lambda^\sigma$ denote the non-dynamical complex parameters which, together with the axio-dilaton $\tau$, parametrise  the non-dynamical $1+b_3(Y)$ marginal deformations of 
the background.

The chiral fields $\Re\rho_a$ and $\Im\rho_a$  provide an alternative parametrisation of the K\"ahler moduli $v^a$ and the $C_4$ moduli, respectively. At the moment, we just need the explicit parametrisation of the $\Re\rho_a$:
\be\label{rhoa0}
\Re\rho_a=\frac12 \sum_I \kappa_a(z_I,\bar z_I;v)-\frac1{2\Im\tau}\,\cali_{a\alpha\beta}\,\Im\beta^\alpha\Im\beta^\beta-\frac1{\Im\tau}\,\cali_{a\alpha\sigma}\,\Im\beta^\alpha\Im\lambda^\sigma\,,
\ee
where we have introduced the intersection numbers $\cali_{a\alpha\beta}=D_a\cdot \hat D_\alpha\cdot \hat D_\beta$, $\cali_{a\alpha\beta}=D_a\cdot \hat D_\alpha\cdot \tilde D_\sigma$, which admit the integral representation 
\be
\cali_{a\alpha\beta}\equiv \int_X\omega_a\wedge \hat\omega_\alpha\wedge\hat\omega_\beta\,,\quad
\cali_{a\alpha\sigma}\equiv \int_X\omega_a\wedge \hat\omega_\alpha\wedge\tilde\omega_\sigma\,.
\ee
By using the asymptotic behaviours (\ref{asymphatomega}) and (\ref{asymphatomega2}), one can indeed check that the above integrals are finite. 
Notice that, as already remarked above, the potentials $\kappa_a(z_I,\bar z_I;v)$ are defined only up to an additive constant, and so is $\Re\rho_a$. The same is true for $\Im\rho_a$, which can be roughly identified with the expansion coefficients  of a flat variation of $C_4$  in a basis of $b_2(X)$ four-forms. These forms are dual, in some appropriate sense, to the harmonic two-forms $\omega_a$.  Their precise definition is complicated by  the presence of the non-trivial self-dual field-strength $F_5$, but fortunately we will not need it in the following. A more explicit description of $\Im\rho_a$ can be found in Appendix \ref{sec:CSterm}.

To explicitly see that $\Re\rho_a$ provide an alternative parametrisation of the K\"ahler moduli, we now show that the relations (\ref{rhoa0})  can be in principle inverted into relations expressing  $v^a$ as functions of $\Re\rho_a,\Im\beta^\alpha,z^i_I$. Indeed, by using (\ref{intdelkappa}) and  (\ref{warpsol}), together with the symmetry of the Green's function, we obtain
\be\label{derrhov}
\begin{aligned}
\frac{\del\Re\rho_a}{\del v^b}=&\frac12\sum_I\frac{\del\kappa_a(z_I,\bar z_I;v)}{\del v^b}=\sum_I\int_{X}G(x_I;x)(J\wedge \omega_a\wedge \omega_b)(x)\\
=&\int_Xe^{-4A}J\wedge \omega_a\wedge \omega_b\equiv -\calg_{ab}\,,
\end{aligned}
\ee
where the matrix $\calg_{ab}$ has been defined in (\ref{defcalg}). Since it  is finite and non-degenerate, (\ref{derrhov})  shows that one can  invert the relations (\ref{rhoa0}).

%%%%%%%%%%%%%%%%%%%%%%%%%%%%%%%%%%%%%%%%%%%%%%%%%%
%%%%%%%%%%%%%%%%%%%%%%%%%%%%%%%%%%%%%%%%%%%%%%%%%%%

\subsection{Effective action and K\"ahler potential}

We are now ready to discuss the low-energy effective theory. Let us assume  that all the D3-branes in the bulk are not mutually coincident and furthermore that the K\"ahler metric on $X$ is smooth enough to justify the validity of the two-derivative  ten-dimensional IIB supergravity.\footnote{In fact, the warping becomes very curved close to the isolated D3-branes, which would suggest a breaking of the leading ten-dimensional supergravity approximation. However, such local geometry is well approximated by a highly curved maximally supersymmetric AdS$_5\times S^5$ background, which is known to be an exact solution of string theory \cite{Berkovits:2004xu}. This suggests that the two-derivative supergravity approximation may be used, for our purposes, beyond its most naive regime of validity, and we will be working with this implicit assumption. See Section \ref{sec:conclusions} for more comments on this point. }

The effective action can be obtained from the rigid/decompactification limit of the supergravity action derived in \cite{Martucci:2014ska} -- see appendix \ref{app:rigidlimit}. One can then write the HEFT Lagrangian as 
\be\label{efflagr}
\call=\call_{\rm YM}+\call_{\rm chiral}\,,
\ee
where 
\be
\call_{\rm YM}= -\frac{1}{4\pi}\sum_{A=1}^{N}\left( \Im\tau F^A \wedge * F^{A}+\Re\tau  F^A \wedge  F^A\right) +\text{(fermions)}\label{YMcontr}
\ee
describes the (trivial) dynamics of $N$ fully decoupled  $U(1)$ SYM theories, while 
\be
\begin{aligned}
\call_{\rm chiral}=&-\pi \calg^{ab} \nabla \rho_a\wedge *\nabla \bar\rho_b-2\pi\sum_I g_{i\bar\jmath}(z_I,\bar z_I)\d z_I^i\wedge *\d \bar z_I^{\bar\jmath}\\
&-\frac{\pi}{\Im\tau}\calm_{\alpha\beta}\d\beta^\alpha\wedge*\d\bar\beta^\beta+\text{(fermions)}\,\label{rigidkineticaxion0}
\end{aligned}
\ee
describes  the (non-trivial) dynamics of the moduli and of their supersymmetric partners. In (\ref{rigidkineticaxion0}), $g_{i\bar\jmath}(z,\bar z)$ is the K\"ahler metric on $X$ and $\calg^{ab}$ is the inverse of the matrix $\calg_{ab}$  introduced in (\ref{defcalg}).
We have also introduced the covariant derivatives $\nabla\rho_a$ and the matrix $\calm_{\alpha\beta}$ defined as follows
\begin{subequations}
\begin{align}
\nabla\rho_a&\equiv\d\rho_a-\cala^{I}_{ai}\d z^i_I-\frac{\ii}{\Im\tau} (\cali_{a\alpha\beta}\Im\beta^\beta+\cali_{a\alpha\sigma}\Im\lambda^\sigma)\d \beta^\alpha\,,\\
\calm_{\alpha\beta}&\equiv \int_X \hat\omega_\alpha\wedge*\hat\omega_\beta= -\int_X J\wedge \hat\omega_\alpha\wedge\hat\omega_\beta=-v^a\cali_{a\alpha\beta}\,,
\end{align}
\end{subequations}
where 
\be\label{defcala}
\cala^{I}_{ai}\equiv \frac{\del\kappa_a(z_I,\bar z_I;v)}{\del z^i_I}   \,.
\ee
The kinetic matrices $\calg^{ab}$ and $\calm_{\alpha\beta}$ are finite exactly because of the conditions (\ref{L2norm}) and (\ref{wL2norm}). 
 Furthermore, note that the kinetic metric for the D3-brane chiral fields $z_I^i$  is the natural covariant extension of the Calabi-Yau metric on $X$. This matches the result obtained by expanding the DBI action of a probe D3-brane and provides a non-trivial consistency check of the validity of our HEFT.
 %- see Section \ref{sec:conclusions} for more comments on this point.}

It remains to show that the effective action (\ref{rigidkineticaxion0}) is consistent with supersymmetry. This is obvious for $\call_{\rm YM}$, 
while it is less trivial to demonstrate that we can write $\call_{\rm chiral}$ in the superspace form 
\be\label{susyform}
\call_{\rm chiral}=\int\d^4\theta K(\Phi,\bar\Phi)=-K_{A\bar B}(\Phi,\bar\Phi) \d\Phi^A\wedge *\d\bar\Phi^{\bar B}+\text{(fermions)}\,,
\ee
with $K_{A\bar B}=\frac{\del^2 K}{\del\Phi^A\del\bar\Phi^{\bar B}}$ for some K\"ahler potential $K(\Phi,\bar\Phi)$, where $\Phi^A$ collectively denote the chiral scalar fields $(\rho_a,\beta^\alpha,z^i_I)$ as well as their complete superfield extension. As we will presently show, such  K\"ahler potential exists and admits the following simple expression in terms of the globally defined function $k_0(z,\bar z;v)$ introduced in section \ref{sec:moduli}:
\be\label{rigidkahler0}
K=2\pi\sum_Ik_0(z_I,\bar z_I;v)\,.
\ee
Notice that this K\"ahler potential is only implicitly defined.  Indeed, it depends on the chiral fields also through the dependence on the K\"ahler moduli $v^a$, which should be expressed as functions of the chiral fields $\rho_a$, $\beta^\alpha$ and $z^i_I$ by inverting (\ref{rhoa0}).

In order to show that (\ref{rigidkahler0}), together with (\ref{rhoa0}),  reproduces (\ref{rigidkineticaxion0}),  we can use (\ref{derrhov}),
which allows us to compute
\be\label{dervrho0}
\frac{\del v^b}{\del\Re\rho_a}=-\calg^{ab}\,,\quad~ \frac{\del v^a}{\del\beta^\alpha}=\frac{\ii}{2\Im\tau}\calg^{ab}\,(\cali_{b\alpha\beta}\Im\beta^\beta+\cali_{b\alpha\sigma}\Im\lambda^\sigma)\,,
\quad~\frac{\del v^a}{\del z^i_I}=\frac12\calg^{ab}\,\cala^{I}_{bi}\,.
\ee
Furthermore, from  (\ref{defk0cond}) and  (\ref{derrhov}), it immediately  follows  that
\be\label{derKv}
\frac{\del K}{\del v^a}=-2\pi v^b\sum_I\frac{\del\kappa_b(z_I,\bar z_I;v)}{\del v^a}=4\pi\,  \calg_{ab}v^b\,.
\ee
From (\ref{rigidkahler0}), (\ref{dervrho0}) and (\ref{derKv})   and taking into account that $K$ depends on $(\rho_a,\bar\rho_a)$ only through $\Re\rho_a$, one can then compute the first derivatives of $K$ with respect to the chiral fields:
\be\label{chiralderK}
\begin{aligned}
\frac{\del K}{\del\rho_a}=&-2\pi v^a\,,\\
\frac{\del K}{\del \beta^\alpha}=&\frac{2\pi\ii}{\Im\tau}\, v^a(\cali_{a\alpha\beta}\Im\beta^\beta+\cali_{a\alpha\sigma}\Im\lambda^\sigma)\,,\\
 \frac{\del K}{\del z^i_I}=&2\pi\Big[v^a\cala^{I}_{ai}+\frac{\del k_0(z_I,\bar z_I;v)}{\del z^i_I}\Big]\,.
 \end{aligned}
\ee
Along the same lines, one can compute the second derivatives $K_{A\bar B}=\frac{\del^2 K}{\del\Phi^A\del\bar\Phi^{\bar B}}$, showing that indeed (\ref{susyform})  reproduces (\ref{rigidkineticaxion0}).

Notice that the HEFT described here does not include possible perturbative as well as non-perturbative string corrections.  We postpone to Section \ref{sec:conclusions} more comments on such corrections. For the moment we just observe that non-perturbative corrections may a priori generate a non-trivial superpotential, which would significantly modify the vacuum structure of the HEFT. In the present setting, such corrections could be generated, if $b_4(X)\neq 0$, by supersymmetric D3-brane instantons. However, as it can be explicitly checked from the complete quadratic fermionic effective action derived \cite{Bianchi:2012kt},  even if supersymmetric  D3-brane instantons existed, they would always carry at least four fermionic zero-modes. This indicates that a non-trivial superpotential is never generated.

The HEFT (\ref{efflagr}) breaks down when two or more D3-branes coincide. Indeed, in this limit the above moduli do not describe the comple light spectrum of the string background, which must include an non-abelian $N=4$ SYM sector. Such break-down  is invisible at the level of the our second-derivative HEFT. This is consistent with the non-renormalisation theorem for $N = 4$ super-Yang-Mills, 
which well approximates the D3-brane sector of the HEFT when they are very close.    
In any case, at such points the internal geometry has no pathologies and  just develops some larger local AdS$_5\times S^5$ throat, which is an exact string theory background \cite{Berkovits:2004xu} and is holographically dual to the additional $N=4$ SYM sector.

\subsection{Structure of the moduli space}
\label{sec:Mstructure}

In this section we discuss in some more detail the structure of the  moduli  space $\calm_{\rm SUGRA}$ of our models, which provides the target space of the non-linear sigma model defining our HEFT.

The D3-brane positions $z^i_I$ parametrise the space 
\be\label{D3moduli}
\calm_{\rm D3}=\text{Sym}^NX\,,
\ee
 while the chiral moduli $\beta^\alpha$ parametrise a $2b_4(X)$-dimensional torus $\calm_\beta$. The additional  $b_2(X)$ chiral coordinates $\rho_a$ (or rather $e^{2\pi\rho_a}$) parametrise the fibres of non-trivial line bundles over $\calm_{\rm D3}\times \calm_\beta$. Hence, $\calm_{\rm SUGRA}$ can be locally identified with the total space of the direct sum of such line bundles. This is most easily seen from the K\"ahler metric on the moduli space, which can be read from the HEFT (\ref{efflagr}):
\be\label{Mmetric}
\d s^2_{\calm_{\rm SUGRA}}=\pi \calg^{ab} {\rm D}\rho_a {\rm D}\bar\rho_b+2\pi\sum_I g_{i\bar\jmath}\,\d z_I^i\d \bar z_I^{\bar\jmath}+\frac{\pi}{\Im\tau}\calm_{\alpha\beta}\d\beta^\alpha\d\bar\beta^\beta\,,
\ee 
where the fibration structure of  $e^{2\pi\rho_a}$ over $\calm_{\rm D3}\times \calm_\beta$ is encoded in the covariant exterior derivative
\be\label{rhocder}
{\rm D}\rho_a =\d\rho_a-\cala^{I}_{ai}\d z^i_I-\frac{\ii}{\Im\tau} (\cali_{a\alpha\beta}\Im\beta^\beta+\cali_{a\alpha\sigma}\Im\lambda^\sigma)\d \beta^\alpha\,.
\ee

In order to better understand the global structure of $\calm_{\rm SUGRA}$, it is convenient to parametrise it in a different way. We first isolate the angular variables  $\phi_a\equiv\Im\rho_a$ and $c^\alpha\equiv\Re\beta^\alpha$, which we collectively denote by $\varphi_\cali$.  They parametrise a $b_2(X)+b_4(X)=\chi(X)-1$ dimensional torus $U(1)^{\chi-1}$ describing the R-R flat potentials.\footnote{The periodicities of the angles $\varphi_\cali$ are determined by the periodicities of the R-R potentials, which are  affected by subtle K-theory corrections \cite{Moore:1999gb}, see footnote \ref{foot:RRK}. } Together with the D3 brane positions $z^i_I$, the angles $\varphi_\cali$ parametrise a space $\calm_0$. Since  the angular variables  $\phi_a$ are fibered over $\calm_{\rm D3}$, $\calm_0$ can be regarded as a fibration of the torus  $U(1)^{\chi-1}$  over $\calm_{\rm D3}$.  

Then, we  substitute $\Re\rho_a$ and $\Im\beta^\alpha$ with the coordinates $\zeta^\cali\equiv(v^a,\chi_\alpha)$,
defined by  a Legendre transform
\be
v^a\equiv -\frac{1}{4\pi}\frac{\del K}{\del \Re\rho_a}\,,\quad \chi_\alpha\equiv\frac{1}{4\pi}\frac{\del K}{\del \Im\beta^\alpha}\,,
\ee
where $v^a$ are just the usual K\"ahler moduli, while the new coordinates $\chi_\alpha$ are given by
\be\label{chidef}
\chi_\alpha=-\frac{1}{\Im\tau}\,v^a\left( \cali_{a\alpha\beta}\Im\beta^\beta +\cali_{a\alpha\sigma}\Im\lambda^\sigma  \right)\,.
\ee
The coordinates  $v^a$ parametrise the standard K\"ahler cone of $X$.  On the other hand, under an overall rescaling  $v^a\rightarrow \lambda v^a$, with $\lambda >0$,   we also have $\chi_\alpha\rightarrow \lambda \chi_\alpha$. Hence   $\zeta^\cali=(v^a,\chi_\alpha)$ parametrise a $b_2(X)+b_4(X)$ dimensional  cone $\calk$.\footnote{ More precisely, $\chi_\alpha$ parametrise a $b_4(X)$ dimensional torus, with $v^a$-dependent periodicity $\chi_\alpha\rightarrow \chi_\alpha+v^a \cali_{a\alpha\beta} n^\beta$ inherited from the integral periodicity of the $B_2$-field. This torus degenerates as $v^a\rightarrow 0$ and the cone $\calk$ can be regarded as the result of fibering it over  the  K\"ahler cone.} Actually, the complete moduli space is given by the extension of $\calk$ to a larger cone, for instance by connecting   different K\"ahler cones by flop transitions.  The internal space $X$ is not generically invariant under such transitions and then the supergravity description generically breaks down at the transition walls. 

We arrive at following global description of the supergravity moduli space $\calm_{\rm SUGRA}$:
\be\label{globalM}
\begin{array}{ccccc} U(1)^{\chi(X)-1}&\hookrightarrow  & \calm_0  & \hookrightarrow & \calm_{\rm SUGRA} \\
& & \downarrow &&\downarrow \\
&& \calm_{\rm D3} && \calk
\end{array}
\ee
Clearly, such global  structure is obscured by the use of the chiral coordinates $(\rho_a, \beta^\alpha, z^i_I)$.

In the new coordinates the moduli space metric (\ref{Mmetric}) reads
\be\label{sgmetric2}
\begin{aligned}
\d s^2_{\calm_{\rm SUGRA}}=&\,\pi\calg_{ab}\d v^a\d v^b+\pi\Im\tau\calm^{\alpha\beta} D \chi_\alpha D\chi_\beta\\
& +\pi\calg^{ab} D\phi_a D\phi_b+\frac{\pi}{\Im\tau}\calm_{\alpha\beta}\d c^\alpha\d c^\beta+2\pi\sum_I g_{i\bar\jmath}\,\d z_I^i\d \bar z_I^{\bar\jmath}\,,
\end{aligned}
\ee
where $\calm^{\alpha\beta}$ is the inverse of $\calm_{\alpha\beta}$,
\be
D\chi_\alpha=\d\chi_\alpha+\frac{1}{\Im\tau}\,\left( \cali_{a\alpha\beta}\Im\beta^\beta +\cali_{a\alpha\sigma}\Im\lambda^\sigma  \right)\d v^a\,,
\ee
and $D\phi_a$ is obtained by taking the imaginary part of (\ref{rhocder}).

We can also express such metric in terms of the  potential obtained by Legendre transform of $K$:
\be
\begin{aligned}
F\equiv&\, K+ 4\pi v^a\Re\rho_a-4\pi \chi_\alpha\Im\beta^\alpha \\
=&\,2\pi\sum_Ik(z_I,\bar z_I;v)+\frac{2\pi}{\Im\tau}v^a\cali_{a\alpha\beta}\Im\beta^\alpha\Im\beta^\beta\, ,
\end{aligned}
\ee
where 
\be
k(z,\bar z;v)=k_0(z,\bar z;v)+v^a\kappa_a(z,\bar z;v)
\ee
is a K\"ahler potential of the internal space: $J=\ii\del\delbar k$.

$F$ must be considered as a function of $(\zeta^\cali, z^i_I,\bar z^{\bar \imath}_I)$. In particular, $\Im\beta^\alpha$ must be considered as functions of $(v^a,\chi_\alpha)$, whose explicit form can be obtained by inverting (\ref{chidef}).
By using the collective coordinates $\zeta^\cali= (v^a,\chi_\alpha)$ and $\varphi_\cali= (\phi_a, c^\alpha)$  the metric (\ref{sgmetric2}) can be rewritten as 
\be\label{sgmetric3}
\d s^2_{\calm_{\rm SUGRA}}=
\,-\frac14 F_{\cali\calj}\d\zeta^\cali\d\zeta^\calj-4\pi^2F^{\cali\calj}\cald\varphi_\cali\cald\varphi_\calj+F^{IJ}_{i\bar\jmath}\d z^i_I\d \bar z^{\bar\jmath}_J\,,
\ee
where
\be
F_{\cali\calj}\equiv \frac{\del^2 F}{\del \zeta^\cali\del \zeta^\calj}\,,\quad F^{IJ}_{i\bar\jmath}\equiv \frac{\del^2 F}{\del z_I^i\del \bar z^{\bar\jmath}_J}\equiv 2\pi \delta^{IJ}g_{i\bar\jmath}(z_I,\bar z_I)\,,
\ee
with $F^{\cali\calj}$ being the inverse of $F_{\cali\calj}$, and 
\be
\cald\varphi_\cali\equiv \d\varphi_\cali-\frac{1}{2\pi}\Im\left(\frac{\del^2 F}{\del \zeta^\cali\del z^i_I}\d z^i_I\right)=\left(\d \phi_a-\Im(\cala^{I}_{ai}\d z^i_I)\,, \d c^\alpha\right)\,.
\ee

Furthermore, the K\"ahler form $\ii\del\delbar K$ on $\calm_{\rm SUGRA}$ reads
\be
\begin{aligned}
\ii\del\delbar K&=-2\pi\d\zeta^\cali\wedge \d\varphi_\cali+\d\, \Im\left(\frac{\del F}{\del z^i_I}\d z^i_I\right)\\
&=-2\pi\d\zeta^\cali\wedge \cald\varphi_\cali+\ii F^{IJ}_{i\bar\jmath}\d z^i_I\wedge \d \bar z^{\bar\jmath}_J\,,
\end{aligned}
\ee 
which shows that the coordinates $\zeta^\cali$ can be regarded as   symplectically paired with the angles $\varphi_\cali$
and that $F$ can be interpreted as a mixed complex-symplectic potential.\footnote{The above Legendre transform can be interpreted as a duality transformation between chiral and linear multiplets \cite{Siegel:1979ai}. Indeed, the function $F$ gives the HEFT in terms of linear multiplets. The linear multiplets are described by real superfields $L^\cali$, such that $\bar D^2 L^\cali=D^2 L^\cali=0$.  Each $L^\cali$ contains  the scalar field $\zeta^\cali$, as lowest component, and a three-form field-strength $\calh^\cali$ which is dual to $\d \varphi_\cali$. The HEFT Lagrangian can then be defined as superspace integral $\int \d^4\theta F(L,z,\bar z)$. If $b_4(X)=0$, its bosonic terms are as in (\ref{dualkinetic}), up to replacing the indices $A,B$ with $a,b$.}

 We then obtain two possible descriptions of the geometry of $\calm_{\rm SUGRA}$.  
A mixed complex-symplectic one  and a purely complex one. On the one hand, the complex-symplectic one appears more `natural' since it better exhibits the global structure   (\ref{globalM}) of the moduli space, the potential $F$ is not  implicitly defined as the K\"ahler potential $K$ and may even more easily accommodate world-sheet quantum corrections. In this sense, one may regard $F$ as the fundamental quantity  and derive $K$ as its anti-Legendre transform. On the other hand, as we discuss in the next section, the chiral coordinates of the complex formulation  can be directly related to the vev of the chiral operators of the dual CFT. Furthermore, they naturally couple to D-brane instantons and then they appear more suitable to describe the complete quantum corrected geometry of the moduli space.

%%%%%%%%%%%%%%%%%%%%%%%%%%%%%%%%%%%%%%%%%%%%%%%%%%%%%%%%%%%%%%%%%%%%%%%%%%%%%%%%%%%%%%%%%%%%%%%%%%%%
%%%%%%%%%%%%%%%%%%%%%%%%%%%%%%%%%%%%%%%%%%%%%%%%%%%%%%%%%%%%%%%%%%%%%%%%%%%%%%%%%%%%%%%%%%%%%%%%%%%%

\section{CFT moduli space} \label{sec:CFT}

In this section we compare the description of the supergravity moduli space provided by the HEFT with the expectations for the moduli space  of the dual CFT.

\subsection{Quiver gauge theories}\label{sec:CFTmoduli}
 
In all the known cases, the CFT corresponds to the IR fixed point of a gauge theory describing $N$ D3 branes probing the Calabi-Yau singularity $C(Y)$. This is given by   an ${\cal N}=1$ quiver gauge theory with gauge group
\be\label{Ggroup}
G= \prod_{i=1}^g SU(N)_i\,,
\ee
 chiral fields $\Phi_a$ transforming in the bi-fundamental representation of pairs of $SU(N)$ factors and a certain superpotential $W(\Phi)$. \footnote{We consider adjoint fields as particular cases of bi-fundamentals connecting the same gauge group.} The number $g$ of $SU(N)$ factors correspond to the Euler characteristic of the resolved space $X$:
 \be
g\equiv \chi(X)=1+b_2(X)+b_4(X)\,.
\ee 
The theory admit marginal deformations that are parametrised by the gauge  and superpotential couplings.
Only a part of these couplings are marginal parameters. Geometrically, we always have at least  $1+b_2(Y)$ marginal parameters that correspond to the parameters $\tau$, $\lambda^\sigma$ of the dual string background. Some CFTs have additional marginal deformations, for example the so-called $\beta$-deformation \cite{Lunin:2005jy},  corresponding to string backgrounds where the internal metric is no more a warped Calabi-Yau.

  The moduli space is given by the solutions of the F and D term conditions
\be 
\label{DFgen}
\frac{\partial W}{\partial \Phi_a}  =  0\, ,\qquad D_{\mathfrak{su}(N)_i} (\Phi_a)   = \, 0 \, ,
\ee 
up to gauge equivalence, where $D_ {\mathfrak{su}(N)_i}$, $i=1,\cdots , g$,  is the moment map for the action of the group $SU(N)_i$. The D-term condition can be omitted if we mod  by the complexified gauge group $G_{\mathbb{C}}$.  
As an affine variety, the moduli space can be indeed written as the quotient of the manifold of F-term solutions by the complexified gauge group
\be \label{quiverM}
{\cal M}= \left\{ \frac{\partial W}{\partial \Phi_a} =0 \right\} \sslash G_{\mathbb{C}} = {\rm Spec} \, \left ( \mathbb{C} \left [ \frac{\partial W}{\partial \Phi_a} =0 \right ]^{G_{\mathbb{C}}}\right )\, .
\ee
By definition, the coordinate ring of this affine variety is just the set of gauge invariant chiral operators made with the $\Phi_a$.
The  gauge invariant chiral operators  are then in one-to-one correspondence with the holomorphic functions on the moduli space and provide a complete characterisation of  the moduli space as an affine complex variety. 

In the toric case, there is an explicit algorithm to write the quiver gauge theory from the toric data 
which is discussed in details in \cite{Franco:2005rj, Hanany:2005ss}.

\subsection{The global symmetries}\label{sec:symmetries}

Of particular importance for us are the global symmetries of the CFT. There are few general observations that can be made for any quiver. The CFT is the IR limit
of the theory of $N$ D3 branes and, in this limit, eventual abelian gauge groups decouple.
Indeed, the gauge group
\be\label{extG}
\tilde G= \left ( \prod_{i=1}^g U(N)_i\right )/U(1) \,.
\ee
 on a set of $N$ D3-branes probing the singularity contains various abelian factors.
The $N$ D3-branes decompose into 
$g=1+b_2(X)+b_4(X)$ stacks of  fractional D3-branes, each supporting a $U(N)_i$ gauge group, and the bifundamental fields correspond to the massless states  of open strings  connecting different fractional branes. The diagonal $U(1)$ is always decoupled and can be modded out as in (\ref{extG}). 

One is then left with $b_2(X)+b_4(X)$ $U(1)$ gauge factors in $\tilde G$, only $b_3(Y)$ of which are  non-anomalous. The anomaly of  the remaining $b_2(X)+b_4(X)-b_3(Y)=2b_4(X)$  $U(1)$'s is cancelled  by a St\"uckelberg mechanism. This can be understood geometrically as follows. One may roughly interpret $b_2(X)+b_4(X)$ fractional D-branes as combinations of  D5 and D7 branes wrapping two- and four-cycles of the resolved geometry. The corresponding $U(1)$'s gauge R-R axions associated with the Poincar\'e dual cohomologies $H^4(X,Y;\mathbb{R})$ and $H^2(X,Y;\mathbb{R})$. However, as we have discussed in section \ref{sec:moduli}, while all $b_4(X)$ independent elements of $H^2(X,Y;\mathbb{R})$ admit an $L_2$-normalisable harmonic representative, only a $b_4(X)$-dimensional subspace of  $H^4(X,Y;\mathbb{R})$. Prior to the near-horizon limit, only these $2b_4(X)$  $L_2$-normalisable modes remain dynamical in the four-dimensional  effective theory, and they are exactly the right number to cancel the corresponding gauge anomalies \`a la St\"uckelberg. The remaining $b_3(Y)$ axions, which would be gauged by the non-anomalous $U(1)$'s,  have infinite  kinetic terms and hence decouple in the four-dimensional  low-energy theory.

The moduli space corresponding to a quiver with gauge group (\ref{extG}) is given by
\be 
{\cal M}_{\mbox{mes}}= \left\{ \frac{\partial W}{\partial \Phi_a} =0 \right\} \sslash \tilde G_{\mathbb{C}} \subset {\cal M}\, ,
\ee
and it is a subvariety of the moduli space of the CFT, ${\cal M}$.  ${\cal M}_{\mbox{mes}}$ is usually called the mesonic moduli space of the CFT. 
${\cal M}_{\mbox{mes}}$ is the set of solutions of the equations
\be 
\label{DFgen22}
\frac{\partial W}{\partial \Phi_a}  =0\, ,\qquad D_{\mathfrak{u}(N)_i} (\Phi)   = 0\, ,
\ee
up to gauge equivalence under the extended gauge group $\tilde G$. 
${\cal M}_{\mbox{mes}}$ is expected to describe the motion of the $N$ D3 branes on the Calabi-Yau singularity. The D3 branes on $C(Y)$ are mutually BPS and we can put them  in arbitrary position.  This implies that the mesonic moduli space is given, as an algebraic variety,  by 
\be \label{mesFI}
{\cal M}_{\mbox{mes}}= \mbox{Sym}^N  C(Y)\, 
\ee
and it has dimension $3 N$. It can be parametrised by the D3-brane positions on $Y$.

The resolution parameters of the Calabi-Yau $X$ enter as FI parameters $\xi_i$ for the D3-brane theory (\ref{extG}). 
The moduli space of D3-branes probing  $X$ is given by the solutions of 
\be 
\label{DFgen2}
\frac{\partial W}{\partial \Phi_a}  =0\, ,\qquad D_{\mathfrak{u}(N)_i} (\Phi)   =  \xi_i  \,  \bbone_{N \times N} \, ,
\ee 
up to gauge equivalence under the extended gauge group $\tilde G$.  Since the overall $U(1)$ is decoupled, one actually has  $\sum_i \xi_i =0$. The moduli space
is now $ \mbox{Sym}^N  X$ and it can be still parametrised by  the D3-brane positions $z^i_I$ on $X$ that are away from the blown-up locus. 

In the IR limit all abelian factors in the D3-brane theory  decouple and become global symmetries of the CFT. 
More precisely, the $b_3(Y)$ non-anomalous $U(1)$ factors  decouple at low energy,  being IR free, while the other $2b_4(X)$ $U(1)$ factors  become massive by the St\"uckelberg mechanism. Hence, at low energy, one is left with the gauge group (\ref{Ggroup}) and  $b_3(Y)$ non-anomalous plus $2 b_4(Y)$    anomalous  global $ U(1)$ symmetries. Such $U(1)$  classical global symmetries  are called baryonic.  Simultaneously, the trace part of the D-flatness condition in (\ref{DFgen22}) gets relaxed. One is then left with $D_{\mathfrak{su}(N)_i} (\Phi)   = 0$, which can be written as
\be 
D_{\mathfrak{u}(N)_i} (\Phi)   =  \, \calv_i \bbone_{N \times N}  \, ,
\ee
where\footnote{In other words, in the near-horizon limit the FI parameters $\xi_i$ appearing in (\ref{DFgen2}) must be rescales appropriately and become dynamical. }
\be 
\calv_i= \frac1N{\rm Tr}\, D_{\mathfrak{u}(N)_i} (\Phi) \, .
\ee
The operator $\calv_i$ is the lowest component of the vector multiplet containing the abelian current corresponding the  $i$-th gauge group.  Notice that $\calv_i$ is not part of a chiral multiplet, but it is nevertheless  protected when the associated baryonic $U(1)$ symmetry is not anomalous.

After the near-horizon limit, the above FI parameters $\xi_i$ can be identified with the expectation values of $\calv_i$, 
\be
\xi_i=\langle \calv_i\rangle\,.
\ee
Now the $\xi_i$ can be regarded as part of  the moduli space and, with some abuse of language, we may refer to them as FI moduli.   Since they still satisfy $\sum_i \xi_i =0$, they parametrise $g-1$ real directions in moduli space.   They naturally pair with the $g-1$ Goldstone bosons associated with the baryonic symmetries. Indeed, in a generic point of the moduli space, the bi-fundamental fields $\Phi_a$ have a vev and the abelian global symmetries are spontaneously broken.   More precisely, the $b_3(Y)$ non-anomalous $U(1)$ symmetries are associated with Goldstone bosons, while the anomalous ones are associated with pseudo-Goldstone bosons.

We then see that the total moduli space  ${\cal M}$  has complex dimension
\be\label{dimmod} {\rm dim} {\cal M} = 3 N + g -1\, .\ee
Indeed ${\cal M}$ can be obtained from (\ref{mesFI}), by relaxing the trace of the D-flatness constraints and by omitting the corresponding $U(1)$ gauge identifications. This gives us the $g-1$ extra complex moduli associated with a complex combination of the FI moduli and the Goldstone bosons. Holographically, they  correspond to the  metric, $B_2$, $C_2$ and $C_4$ moduli of the dual resolved Calabi-Yau, which are dynamical in the near-horizon geometry.

 By comparing with Section \ref{sec:EFT},  we can make the identification $\calm\simeq \calm_{\rm SUGRA}$ and set the correspondence with the string theory moduli given in Table \ref{tablemoduli}.
\begin{table}[h!!!!!!!]
\begin{center}
  \begin{tabular}{ | c | c | c | c |}
    \hline
    classical $U(1)$s  & harmonic 2-forms & (pseudo)-Goldstone chiral fields & Betti number \\ \hline\hline
   anomalous   & $\hat\omega_\alpha$  &$\hat\rho_\alpha$ & $b_4(X)$  \\ \hline
   anomalous   & $\hat\omega_\alpha$  &$\beta^\alpha$ & $b_4(X)$ \\ \hline
non-anomalous   & $\tilde\omega_\sigma$  &$\tilde\rho_\sigma$ & $b_3(Y)$ \\ 
    \hline
  \end{tabular}
\end{center}
\caption{(Pseudo) -  Goldstone bosons.}
\label{tablemoduli}
\end{table}

As we noticed in Section \ref{sec:EFT}, we can use complex as well as complex-symplectic coordinates to  describe the  moduli space.
The HEFT chiral fields $\rho_a,\beta^\alpha, z^i_I$ provide a holographic complex parameterisation of the CFT moduli space ${\cal M}$. The vev of a gauge invariant chiral operator  is a holomorphic function on ${\cal M}$ and therefore should be expressible as a holomorphic function of $\rho_a,\beta^\alpha, z^i_I$. 
On the other hand, one may use the alternative  complex-symplectic coordinates $\zeta^\cali,\varphi_\cali, z^i_I$. The variables $z^i_I$ parametrise the motion of the D3-branes on
the resolved cone $X$. The angles $\varphi_\cali=(\phi_a,c^\alpha)$, $\cali=1,\ldots, g-1$, correspond to the baryonic Goldstone and pseudo-Goldstone  (real) bosons, while the symplectic coordinates $\zeta^\cali=(v^a,\chi_\alpha)$ can be set in correspondence with FI moduli $\xi_i$ (taking into account the constraint $\sum_i \xi_i =0$). 

 At the generic point of the moduli space, the CFT microscopic gauge theory group  is spontaneously broken to $N-1$ decoupled $U(1)$ factors, plus the overall diagonal  $U(1)$ of the parent quiver $U(N)$ theory which, being always decoupled, is usually ignored. Then, at low energy, there is a total of $N$ trivial and fully decoupled  SYM $U(1)$ sectors, which are represented by the contribution (\ref{YMcontr}) to the HEFT.

\subsection{Comparison with  the AdS/CFT correspondence}\label{sec:AdSbaryons}

In the spirit of the AdS/CFT correspondence, smooth backgrounds with the same boundary asymptotics describe different  vacua of the same theory. In our case, AdS$_5\times Y$ itself  corresponds to the origin of the moduli space ${\cal M}$. All other vacua in ${\cal M}$ are associated with smooth  backgrounds asymptotic to AdS$_5\times Y$. As we have discussed, they  correspond to the near horizon geometries of D3-branes
moving on the resolved Calabi-Yau. 

The non-anomalous baryonic symmetries are easy to identify in terms of the geometry of $Y$. They are associated with the massless vectors in the bulk  that arise from the reduction of $C_4$ on $Y$. There are precisely $b_3(Y)$ of them.  

In  the AdS/CFT correspondence, the natural objects  to consider   are  the gauge invariant operators.  The corresponding bulk fields arise from the KK spectrum of AdS$_5\times Y$
and from wrapped branes. The chiral KK modes on $Y$ are in one-to-one correspondence with the mesonic operators with zero baryonic charge.
On the other hand, baryonic operators are obtained by wrapping Euclidean D3-branes on non trivial three-cycles $\Sigma\subset Y$. 
A Euclidean D3-brane is supersymmetric when  the complex cone $C(\Sigma)$ is a divisor in the Calabi-Yau cone $C(Y)$. 
The more general classical supersymmetric D3-brane configuration is
obtained by considering arbitrary divisors and it is expected  that all baryonic operators in the CFT arise by a geometric quantisation of these classical configurations \cite{Mikhailov:2000ya,Beasley:2002xv}. We can consider also divisors that are trivial and correspond to contractible three-cycles in $Y$. The interpretation of the corresponding state is in terms of giant gravitons \cite{Beasley:2002xv}.

This point of view is particularly useful because every elementary field $\Phi$ in the quiver transform in the bi-fundamental or adjoint representation of the gauge group $G$
and therefore, by a double determinant,  we can always construct a gauge invariant operator, schematically denoted by $\det \Phi$. There should exist therefore a conical divisor $D$ in $C(Y)$ corresponding to the field $\Phi$.
A D3-brane wrapped on the base of $D$ corresponds to the operator $\det\Phi$. When the base of $D$ in $Y$ is non-trivial, this is a baryonic operator. When the base is trivial, the
operator is equivalent to a complicated linear combination of mesonic operators.\footnote{The standard example is AdS$_5\times S^5$ where all the three-cycles are trivial. The determinant of any elementary adjoint fields $\Phi$ in ${\cal N}=4$ SYM can be written in terms of a linear combinations of product of traces, $ \det\Phi =  \sum_{n_1+\ldots +n_p=N}  c_{n_1\cdots n_p}   {\rm Tr} \Phi^{n_1}\cdots {\rm Tr} \Phi^{n_p}$, using the tensor identity  $\epsilon^{a_1\cdots a_N} \epsilon_{b_1\cdots b_N} = N!\delta^{a_1}_{[b1} \cdots  \delta^{a_N}_{b_N]}$.}
This identification  allows to compute
the dimension of a baryonic operator ${\cal B}$ associated with a divisor $D$ using purely geometrical data as \cite{Gubser:1998fp}
\be\label{volD} \frac{N\pi\text{vol}(\Sigma)}{2\text{vol}(Y)} \, , \ee
where $\Sigma\subset Y$ is the base of $D$.

We then expect, in general,  a correspondence between elementary fields and conical divisors in $C(Y)$.
This correspondence is well understood for  toric Calabi-Yau cones \cite{Franco:2005rj, Benvenuti:2005ja, Butti:2007jv,Feng:2005gw,Hanany:2005ss,Forcella:2008bb}. It allows to reconstruct  the quiver gauge theory form toric data and to compute dimension
and R-charges of the elementary fields in the CFT from geometry.

More interestingly for us,  we can also probe the vev of the baryonic operator associated with a divisor $D$ by evaluating the Euclidean action of a D3-brane wrapping the corresponding divisor in the resolved space $X$ \cite{Klebanov:2007us}. This can be re-interpreted in the language of our HEFT, as it will be discussed in details  in the next Section.

%%%%%%%%%%%%%%%%%%%%%%%%%%%%%%%%%%%%%%%%%%%%%%%%%%%%%%%%%%%%%%%%%%%%%%%%%%%%%%%%%%%%%%%%%%%%%%%%%%%%
%%%%%%%%%%%%%%%%%%%%%%%%%%%%%%%%%%%%%%%%%%%%%%%%%%%%%%%%%%%%%%%%%%%%%%%%%%%%%%%%%%%%%%%%%%%%%%%%%%%%

%%%%%%%%%%%%%%%%%%%%%%%%%%%%%%%%%%%%%%%%%%%%%%%%%%%%%%%%%%%%%%%%%%%%%%%%%%%%%%%%%%%%%%%%%%%%

\section{Baryonic vevs from Euclidean D3-branes}
\label{sec:baryonvev}

In the unbroken phase, the chiral baryonic operators are associated with supersymmetric Euclidean D3-branes (E3-branes) 
wrapping non-compact divisors of the Calabi-Yau cone $C(Y)$ \cite{Mikhailov:2000ya,Beasley:2002xv}. 
This correspondence is assumed to remain true even in the baryonic phase, in which the FI moduli $\xi_i$ are non-vanishing and the internal space $X$ is correspondingly  resolved into a smooth space.\footnote{This correspondence is valid only at the semiclassical level. More precisely, a baryon is associated with a state of the Hilbert space which is obtained by quantising the moduli space of the divisor \cite{Beasley:2002xv}. See later for further discussions on this point. } In particular, a natural subclass of  baryons  is associated with asymptotically conical effective divisors. Hence, according to
the prescription adopted in \cite{Klebanov:2007us} for the KW theory,  the vev of the baryonic operator $\calb$ with dimension $\Delta_\calb$ associated with an asymptotically conical non-compact  divisor $D$  can be extracted from the schematic  semiclassical formula
\be\label{vevbaryons}
e^{-S_{\rm E3}}\simeq r_{\rm c}^{-\Delta_\calb}\,\langle \calb\rangle\,,
\ee
where $ r_c$ represents an UV cut-off that regularises the on-shell action $S_{\rm E3}$. 

The extension of this procedure to more general theories has been discussed in some detail in \cite{Martelli:2007mk,Martelli:2008cm}. In this section we would like to compute   $\langle \calb\rangle$ in terms of our supergravity chiral fields $\rho_a,\beta^\alpha,z^i_I$. We then need to compute $S_{\rm E3}$ associated with a supersymmetric  E3-brane wrapping $D$ and supporting a line-bundle with fixed boundary condition specified as follows. 

We denote by $\Sigma=\del D$ the asymptotic boundary of $D$ and as in \cite{Martelli:2008cm} we assume that $b_1(\Sigma)=0$, $H_1(D;\mathbb{Z})=0$ and $H^2(D;\mathbb{C})\equiv H^{1,1}(D;\mathbb{C})$, which indeed hold for most of the known explicit examples (e.g.\ in the toric cases). We then have $H_1(\Sigma;\mathbb{Z})\equiv H_1(\Sigma;\mathbb{Z})_{\rm tor}$ and we can write the short exact sequence 
\be\label{splitDcycles}
0\longrightarrow H_2(D;\mathbb{Z})\longrightarrow H_2(D,\Sigma;\mathbb{Z})\longrightarrow H_1(\Sigma;\mathbb{Z})_{\rm tor}\longrightarrow 0\,.
\ee
Now, a line bundle $L$ on $D$ is associated with a certain element of $H^2(D;\mathbb{Z})$ by its first Chern class. By Poincar\'e duality, we can regard it as a relative two-cycle in $H_2(D,\Sigma;\mathbb{Z})$, which can have a torsional one-cycle  $\gamma\subset \Sigma$ as boundary. Hence, fixing the boundary condition for the allowed line bundle $L$ 
corresponds to fixing such torsional one-cycle  $\gamma\subset \Sigma$,  which in turn corresponds to fixing the torsional line bundle  $L|_\Sigma$ on the boundary $\Sigma$. On the other hand
from (\ref{splitDcycles}) it is clear that there are different  line bundles with the same fixed boundary condition. They are counted by  the two-cycles in  $H_2(D;\mathbb{Z})$, which are  Poincar\'e dual to compactly supported world-volume fluxes in  $H^2(D,\Sigma;\mathbb{Z})$. 
Therefore, on the r.h.s\ of  (\ref{vevbaryons}) one should actually sum over all line bundles on $D$ which define the same flat torsional line bundle on the boundary $\Sigma$.  

Let us denote by $F$ the world-volume flux associated with the line bundle $L$ plus the possibile half-integer shift due to the Freed-Witten anomaly, so that $\frac1{2\pi}[F]=c_1(L)+\frac12 c_1(D)$ \cite{Minasian:1997mm,oai:arXiv.org:hep-th/9907189}. This then naturally combines with the $B_2$ field into the gauge invariant field-strength 
\be
\calf\equiv \frac{\ell^2_{\rm s}}{2\pi}F-B_2|_D\,.
\ee
A detailed discussion of the supersymmetry of Euclidean D-branes in $N=1$ backgrounds can be found in \cite{eff1}, and the resulting conditions can be expressed in terms of the generalised calibrations of \cite{Koerber:2005qi,luca1}. In our setting, they traslate into the condition that the E3-brane warps a holomorphic submanifold, as we are assuming, and that $\calf$ is anti-self-dual:
\be\label{ASDcalf}
*_D\calf=-\calf\,. 
\ee

In \cite{Martelli:2008cm} it is argued that, under our topological assumptions, $H^2(D,\Sigma;\mathbb{R})\simeq H^2(D;\mathbb{R})\simeq \calh_{L_2}^2(D)$, so that any element of $H^2(D;\mathbb{R})$ admits an  harmonic representative, which is $L_2$-normalisable with the respect to the metric induced on $D$. In particular, we can choose a  basis  of   $L_2$-normalisable harmonic $(1,1)$-forms $\alpha_k$, $k=1,\ldots,b_2(D)$, and  a basis of two-cycles $C^k\subset D$ in $H_2(D;\mathbb{Z})$  such that $\int_{C^k}\alpha_l=\delta^k_l$. We can then define
\be
f^k\equiv \frac1{2\pi}\int_{C^k}F\,,\quad \hat N^k{}_\alpha\equiv \int_{C^k}\hat\omega_\alpha\,,\quad \tilde N^k{}_\sigma\equiv \int_{C^k}\tilde\omega_\sigma\,,\quad \cali^D_{kl}\equiv \int_D\alpha_k\wedge\alpha_l\,.
\ee
Notice that $\hat N^k{}_\alpha,\tilde N^k{}_\sigma\in\mathbb{Z}$ while $\cali^D_{kl}$ is a negative definite symmetric matrix which may not be integrally quantised.\footnote{\label{foot:posID} One can see that $\cali^D_{kl}$ is negative definite by rewriting it as $-\int_D\alpha_k\wedge*\alpha_l$. This is possible since, by following the same argument used  for the bulk (1,1)-forms $\omega_a$,  one can show that also the (1,1)-forms $\alpha_k$ on $D$ are primitive and then anti-self-dual. We then observe that, as in \cite{Martelli:2008cm},  one can write $\int_D\alpha_k\wedge\alpha_l=[\alpha_k]_{\rm cpt}\cup [\alpha_l]$, where  $[\alpha_k]_{\rm cpt}$ is a representative of $\alpha_k$ in $H^2(D,\Sigma;\mathbb{R})$. On the other hand, from the short exact sequence Poincar\'e dual to (\ref{splitDcycles})
%\be\label{integerDseq}
%0\longrightarrow H^2(D,\Sigma;\mathbb{Z})\longrightarrow H^2(D;\mathbb{Z})\longrightarrow {\rm Tor}H^2(\Sigma;\mathbb{Z})\longrightarrow 0\,.
%\ee
we see that, since $[\alpha_k]\in H^2(D;\mathbb{Z})$ and  $H^2(\Sigma;\mathbb{Z})_{\rm tor}$ can be non-trivial, $[\alpha_k]_{\rm cpt}$ does  not necessarily define an element of $H^2(D,\Sigma;\mathbb{Z})$. Rather, we can always choose a minimal $n_k\in \mathbb{Z}$ such that $[n_k\alpha_k]_{\rm cpt} \in H^2(D,\Sigma;\mathbb{Z})\simeq H_2(D;\mathbb{Z})$. Hence, in general the entries of $\cali_{kl}$ are just rational.}

Being $\calf$ closed and anti-self-dual, it is an $L_2$-normalisable harmonic form. Hence it can be expanded as follows
\be\label{calfexp}
\begin{aligned}
\calf&=\alpha_k \int_{C^k}\calf= \ell^2_{\rm s}\,\alpha_k\left[f^k+\frac{1}{\Im\tau}(\hat N^k{}_\alpha\Im\beta^\alpha+\tilde N^k{}_\sigma\Im\lambda^\sigma)\right]\,.
\end{aligned}
\ee 
 Furthermore, from the above discussion it follows that we can expand $\frac{1}{2\pi}[F]\in H^2(D;\mathbb{Z})\simeq H_2(D,\Sigma;\mathbb{Z})$ as follows:
\be
\frac{1}{2\pi}[F]=\frac12\calc_D+\tilde C_0+m_l C^l\,,
\ee
 with $m_l\in \mathbb{Z}$, $\calc_D\equiv [c_1(D)]$ and $\tilde C_0$ any fixed element of $H_2(D,\Sigma;\mathbb{Z})$ representing a line bundle with the appropriate boundary conditions. We can then write 
\be
f^k=f^k_0 + m_lM^{kl}\,,
\ee
where $M^{kl}\equiv C^k\cdot C^l$ and $f^k_0\equiv C^k\cdot(\tilde C_0+\frac12\calc_D)$. \footnote{In may be convenient to minimise $f^k_0$ by redefining  $\tilde C_0\rightarrow \tilde C_0+n_l C^l$ for some $n_l\in\mathbb{Z}$.} Hence the vector ${\bf f}=(f^1, \ldots , f^{b_2(D)})$ takes value in the shifted lattice ${\bf f}_0+{\bf M}\mathbb{Z}^{b_2(D)}$. Notice that $M^{kl}$ is the inverse of $\cali^D_{kl}$ and furthermore  $[C^k]^{\rm h}=M^{kl}\alpha_l$, where $[C^k]^{\rm h}$ is the harmonic representative of the Poincar\'e dual of $C^k$.

\subsection{DBI contribution}
\label{sec:DBIcontr}

The calibration condition implies that the on-shell DBI action can be written as
\be\label{warped4Dvolume}
\frac{1}{2\pi}S^{\rm DBI}_{\rm E3}(D)=\frac12\int_{D} e^{-4A} J\wedge J -\frac1{2\ell^{4}_{\rm s}}{\Im\tau}\int_{D} \calf\wedge \calf-\frac1{24}\Im\tau\,\chi(D)\,,
\ee
where $\chi(D)\equiv \int_D c_2(D)$ is the Euler characteristic of the divisor $D$, which has been introduced by supersymmetrisation of  the curvature correction \cite{Green:1996dd,Cheung:1997az,Minasian:1997mm} to the CS action, see appendix \ref{sec:CSterm}.
We would like to express (\ref{warped4Dvolume}) in terms of our background moduli and parameters. 

Let us start with the second term on the r.h.s.\ of (\ref{warped4Dvolume}).
By expanding $[D]=n^a [D_a]$ in $H_4(X,Y;\mathbb{Z})$, where $D_a$ are a basis of divisors Poincar\'e dual to the bulk harmonic forms $\omega_a$, we can derive the identities
\be\label{identities}
\cali^D_{kl}\,\hat N^k{}_\alpha\hat N^l{}_\beta=n^a\cali_{a\alpha\beta}\equiv \cali^D_{\alpha\beta}\,,\quad \cali^D_{kl}\,\hat N^k{}_\alpha\tilde N^l{}_\sigma=n^a\cali_{a\alpha\sigma}\equiv\cali^D_{\alpha\sigma}\,. 
\ee
Similarly, we define
\be\label{noncompactint}
\cali^D_{\sigma\rho}\equiv \cali^D_{kl} \tilde N^k{}_\sigma\tilde N^l{}_\rho\,.
\ee

By using the expansions (\ref{C2B2exp}) and (\ref{calfexp}),  we can now rewrite the second term on the r.h.s.\ of (\ref{warped4Dvolume}) as follows 
\be
\begin{aligned}
-\frac1{2\ell^{4}_{\rm s}}{\Im\tau}\int_{D} \calf\wedge \calf\equiv& -\frac1{2\Im\tau} \cali^D_{\alpha\beta}\Im\beta^\alpha\Im\beta^\beta -\frac1{\Im\tau} \cali^D_{\alpha\sigma}\Im\beta^\alpha\Im\lambda^\sigma\\
& -\frac1{2\Im\tau}\cali^D_{\rho\sigma}\Im\lambda^\sigma\Im\lambda^\rho +I^{\rm DBI}_F({\bf f})\,,
\end{aligned}
\ee
 with
\be
I^{\rm DBI}_F({\bf f})=-\cali^D_{kl}(\hat N^k{}_\alpha \Im\beta^\alpha+
\tilde N^k{}_\sigma \Im\lambda^\sigma)f^l-\frac12\Im\tau\cali^D_{kl} f^kf^l\,.
\ee

We can now pass to the first term on  the r.h.s.\ of (\ref{warped4Dvolume}).
As in  \cite{Martucci:2014ska}   (see also \cite{Benishti:2010jn} for an analogous argument in the relative M-theory context), 
it is convenient to rewrite it as follows:
\be\label{ww4}
\frac12\int_{D} e^{-4A} J\wedge J=\frac12\int_X e^{-4A} J\wedge J\wedge\delta^2(D)=\int_X e^{-4A} J\lrcorner \delta^2(D)\d\text{\rm vol}_X\,.
\ee
We can then use the identity $\delta^2(D)=\frac{1}{2\pi}\ii\del\delbar\log|\zeta_D(z)|^2$, where $\zeta_D(z)$ is a non-trivial section of $\calo(D)$ defining the divisor $D=\{\zeta_D(z)=0\}$, which implies that
\be\label{PL}
J\lrcorner \delta^2(D)=-\frac1{4\pi}\Delta\log|\zeta_D|^2\,.
\ee
In order to make this formula useful notice that, since the harmonic (1,1) form $\omega_D\equiv n^a\omega_a$ is primitive, the associated locally defined potential $\kappa_D\equiv n^a\kappa_a$ (such that $\omega_D=\ii\del\delbar\kappa_D$)  is harmonic: $\Delta\kappa_D=0$. Then we can actually write
\be\label{JD}
J\lrcorner \delta^2(D)=-\frac1{4\pi}\Delta h_D\,,
\ee
where 
\be
h_D(x)\equiv \log\left(|\zeta_D|^2e^{-2\pi\kappa_D}\right)(x)\,.
\ee 
This function is nothing but the norm of the holomorphic section $\zeta_D$ and (\ref{JD}) tells us that we can regard  $h_D$ as a harmonic  function on $X\backslash D$ which is `sourced' by the divisor $D$.

The advantage of modifying (\ref{PL}) in this way is that $h_D$ is globally defined, 
while $\log|\zeta_D|^2$ is only locally defined.
Then, we can substitute it in (\ref{ww4}) and integrate by parts twice, getting
\be\label{ww5}
\begin{aligned}
\frac12\int_{D} e^{-4A} J\wedge J=&-\frac{1}{4\pi}\int_{X}h_D \Delta e^{-4A}\d\text{\rm vol}_X+I_{ \Sigma}(r_{\rm c})
\\
&=\frac12 n^a\sum_I \kappa_a(z_I,\bar z_I;v)-\frac1{2\pi} \sum_I \Re\log\zeta_D(z_I)+I_{\Sigma}(r_{\rm c})\,,
\end{aligned}
\ee
where the boundary contribution $I_{ \Sigma}(r_{\rm c})$ is given by 
\be\label{boundaryterm}
\begin{aligned}
I_{ \Sigma}(r_{\rm c})=&\frac1{4\pi}\int_{Y,r_{\rm c}}\Big(e^{-4A}*_X\d h_D-h_D*_X\d e^{-4A} \Big)\\
%=&\frac1{2}\lim_{r_{\rm c}\rightarrow \infty}r^5_{\rm c}\int_{Y_{{\rm c}}}\d\text{vol}_Y\Big(e^{-4A}\del_r f^a-f^a\del_r e^{-4A} \Big)=\\
=&\frac1{4\pi}R^4\left( r_{\rm c}\int_{Y,r_{\rm c}}\d\text{vol}_Y \del_r h_D+4\int_{Y,r_{\rm c}}\d\text{vol}_Y h_D \right)\,,
\end{aligned}
\ee
and $r_{\rm c}$ is a very large UV regulator, which will be eventually sent to $\infty$. We have implicitly used the asymptotic warping (\ref{asymwarp}) and the fact that the five-dimensional manifold defined by $\{r=r_c\}$  coincides with the Sasaki-Einstein manifold $Y$ in the  $r_{\rm c}\rightarrow \infty$ limit.  

In order to compute the behaviour of $I_{ \Sigma}(r_{\rm c})$ for $r_{\rm c}\rightarrow\infty$, we just need  the behaviour of $h_D(x)$ at the boundary $r\simeq r_{\rm c}$. In this region the metric is well approximated by the conical one (\ref{conmetric}). One can then use the expansion $ h_D(x)=  h_D^\lambda(r)\alpha_\lambda(y)$  in an orthogonal basis $\alpha_\lambda$ of eigenfunctions of the Sasaki-Einstein 
Laplace operator on $Y$ such that $\Delta_Y\alpha_\lambda=\lambda\alpha_\lambda$, with $\lambda\geq 0$. In particular, we can choose $\alpha_0(y)\equiv 1$ as zero-mode. Clearly, only such zero-mode contributes to (\ref{boundaryterm}), which, using also (\ref{radius}),  becomes
\be\label{boundaryterm2}
\begin{aligned}
I_{ \Sigma}(r_{\rm c})
=&\frac1{16\pi}N\left[ r_{\rm c}\del_{r_{\rm c}} h_D^0(r_{\rm c})+4 h_D^0(r_{\rm c}) \right]\,.
\end{aligned}
\ee
It then remains to solve  the equation for $h_D^0$ obtained by expanding (\ref{JD}), which is given by
\be\label{eqf0}
\frac{1}{r^5}\frac{\d}{\d r}\Big(r^5\frac{\d h_D^0(r)}{\d r}\Big)|_{r=r_{\rm c}} = \frac{4\pi}{\text{vol}(Y)}\calj(D,r_{\rm c}) \quad~~~~~~~\text{(for $r_{\rm c}\rightarrow \infty$)}\,,
\ee
where we have used the asymptotic form $\Delta_{\rm cone}=-\frac{1}{r^5}\del_r(r^5\del_r)+\frac{1}{r^2}\Delta_Y $ of the Laplace operator and we have introduced the quantity
\be
\calj(D,r_{\rm c})=\int_{Y,r_{\rm c}}\d\text{vol}_{Y} J\lrcorner \delta^2(D) \,.
\ee
Since $D$ is asymptotically conical, we can use the formula derived in Appendix \ref{app:asympt}  and write
\be
\calj(D,r_{\rm c})\simeq\frac{1}{r^2_{\rm c}}\text{vol}(\Sigma)\,.
\ee
Equation  (\ref{eqf0})  is readily integrated into
\be
h_D^0(r_{\rm c})\simeq c+\frac{\pi\text{vol}(\Sigma)}{\text{vol}(Y)} \log r_{\rm c}\,.
\ee
By using such asymptotic expansion in (\ref{boundaryterm2}) we arrive at
\be\label{SE3asympt}
I_{ \Sigma}(r_{\rm c})\simeq\frac{N\text{vol}(\Sigma)}{4\text{vol}(Y)} \log r_{\rm c}\,,
\ee
up to an additive constant, which can be reabsorbed by a rescaling of the holomorphic section $\zeta_D$.

We conclude that $2\pi I_{ \Sigma}(r_{\rm c})$ provides the only (logarithmically) divergent contribution to $S^{\rm DBI}_{\rm E3}$. By comparing (\ref{SE3asympt}) with (\ref{vevbaryons}), we arrive at the identification
\be
\Delta_\calb\equiv \frac{N\pi\text{vol}(\Sigma)}{2\text{vol}(Y)}\,,
\ee
which is indeed the expected dimension of the baryon $\calb$, see (\ref{volD}).  Hence,  as already argued in \cite{Martelli:2007mk}, the DBI action has the correct divergent contribution to match (\ref{vevbaryons}). 

We can now combine the different pieces to write the DBI-action in function of our background chiral moduli. Indeed, recalling (\ref{rhoa0}) and the definitions (\ref{identities}), we can write
\be\label{DBIcontr}
\begin{aligned}
S^{\rm DBI}_{\rm E3}=&\,2\pi n^a\Re\rho_a- \sum_I \Re\log\zeta_D(z_I) +\log r^{\Delta_\calb}_{\rm c}\\
&-\frac\pi{\Im\tau}\cali^D_{\rho\sigma}\Im\lambda^\sigma\Im\lambda^\rho-\frac{\pi}{12} \Im\tau\, \chi(D)+2\pi I^{\rm DBI}_F({\bf f})\,.
\end{aligned}
\ee

\subsection{Complete E3-brane action and baryonic vev}
 
The complete E3 on-shell action is given $S_{\rm E3}=S_{\rm E3}^{\rm DBI}+\ii S^{\rm CS}_{\rm E3}$. 
The CS contribution is slightly more subtle than the DBI term and is discussed in some detail in appendix \ref{sec:CSterm}.
The bottom line is that $S_{\rm E3}$ is given by the following  natural completion of  (\ref{DBIcontr})
\be
\begin{aligned}
S_{\rm E3}({\bf f})=&\,2\pi n^a\rho_a -\sum_I\log\zeta_D(z_I)+\log r^{\Delta_\calb}_{\rm c}+2\pi I_F({\bf f})+2\pi\ii c(\tau,\lambda) \,,
\end{aligned}
\ee
with
\be
\begin{aligned}
I_F({\bf f})\equiv&\, \ii\cali^D_{lk}(\hat N^k{}_\alpha \beta^\alpha+   \tilde N^k{}_\sigma \lambda^\sigma)f^l+\frac{\ii}{2}\tau \cali^D_{kl} f^k f^l\,,\\
c(\tau,\lambda)\equiv&\,\frac{1}{24}\tau\,\chi(D) +\frac{1}{2\Im\tau}\cali^D_{\sigma\rho} \lambda^\sigma\Im\lambda^\rho\,.
\end{aligned} 
\ee

We can finally compute the baryonic vev. As discussed above, the relation (\ref{vevbaryons}) must be modified into
\be\label{vevbaryons1}
 \langle \calb\rangle =r_{\rm c}^{\Delta_\calb} \sum_{{\bf f}\in {\bf f}_0+{\bf M}\mathbb{Z}^{b_2(D)}}e^{-S_{\rm E3}({\bf f})}\,,
\ee
where ${\bf M}$ represents the intersection matrix $M^{kl}=C^k\cdot C^l$. This gives
\be\label{expVEV}
\langle\calb\rangle=e^{-2\pi\ii c(\tau,\lambda)}\cala(\beta)\prod_{I}\zeta_D(z_I)  e^{-2\pi n^a\rho_a}\,,
\ee 
where
\be
\begin{aligned}
\cala(\beta)&\equiv \sum_{{\bf f}\in {\bf f}_0+{\bf M}\mathbb{Z}^{b_2(D)}}e^{-2\pi I_F({\bf f})}=\Theta\left[\begin{array}{c} \cali^D{\bf f}_0 \\  0 \end{array}\right]\Big(-\hat {\bf N}{\bf \beta}-\tilde{\bf N}\lambda\Big|-\tau {\bf M}\Big)\,.
\end{aligned}
\ee
Here we are using an obvious matrix notation and $\Theta\left[\begin{array}{c}  {\bf a} \\ {\bf b} \end{array}\right]({\bf w}| {\bf T})$ is the theta function with characteristics $({\bf a},{\bf b})$:
\be
\Theta\left[\begin{array}{c}  {\bf a} \\ {\bf b} \end{array}\right]({\bf w}| {\bf T})=\sum_{{\bf n}\in \mathbb{Z}^{b_2(D)}}\exp\left\{2\pi\ii \left[({\bf n}+{\bf a})_k ({\bf w}+{\bf b})^k+\frac12({\bf n}+{\bf a})_k T^{kl} ({\bf n}+{\bf a})_l\right]\right\}\,.
\ee
Since the matrix $\cali^D_{kl}$ is negative definite, see footnote \ref{foot:posID}, and $M^{kl}$ is its inverse,  then
the matrix $\Im(-\tau {\bf M})$ is positive definite and the theta function is well defined.

We see that, up to a constant, $\langle\calb\rangle$ is completely determined by the chiral fields entering the HEFT  in  a manifestly holomorphic way, which is indeed one of its expected properties. The appearance of a theta function depending on the $B_2$ and $C_2$ moduli in this type of evaluation of the baryonic vev was already pointed out in \cite{Martelli:2008cm} and is expected from the general discussion of  \cite{Witten:1996hc} for the dual five-brane. Here we have made manifest its compatibility with structure of our HEFT. The proper understanding of the global properties of this holomorphic dependence would require a better study of the K-theory corrections to the R-R periodicities, see footnote \ref{foot:RRK}, which will not be addressed in the present paper.

We remark that (\ref{expVEV}) gives just a semiclassical formula for the baryonic condensates. In fact, in order to obtain a more precise identification of the corresponding baryonic operators, one must quantise the E3-brane moduli space, as in \cite{Beasley:2002xv}. This means that one must consider $\langle \calb\rangle$ in     (\ref{expVEV}) as a section of an appropriate line bundle $\call_{\calb}$ over the moduli space of the divisor $D$.  This holomorphic section must be considered as a wave-function in the Hilbert space of BPS E3-branes, which can be expanded in a basis of global sections of $\call_{\calb}$, corresponding to a basis of baryonic operators. The coefficients of this expansion can be then identified with the vev of the corresponding operators. See  Section \ref{sec:KWbaryons} for an explicit illustration of this procedure for the Klebanov-Witten model.

%%%%%%%%%%%%%%%%%%%%%%%%%%%%%%%%%%%%%%%%%%%%%%%%%%%%%%%%%%%%%%%%%%%%%%%%%%%%%%%%%%%%%%%%%%%%%%%%%%%%
%%%%%%%%%%%%%%%%%%%%%%%%%%%%%%%%%%%%%%%%%%%%%%%%%%%%%%%%%%%%%%%%%%%%%%%%%%%%%%%%%%%%%%%%%%%%%%%%%%%%

\section{The HEFT of the Klebanov-Witten theory}
\label{sec:conifold}

In this section we focus on the KW model \cite{Klebanov:1998hh}, presenting a detailed  discussion of its HEFT. This   will  illustrate how to concretely apply    our general results in  a prototypical example. It would be interesting to extend this analysis to other models, in particular to understand some aspects, like the anomalous $U(1)$ symmetries,  which are not present in the KW model.

\subsection{CFT of the KW model}

Let us start by briefly reviewing the structure of the CFT of the KW model and its moduli space from a field theoretical perspective.

The field theory describe $N$ D3-branes probing the singular conifold \cite{Candelas:1989js}. 
The gauge group is $SU(N)\times SU(N)$, there are four bi-fundamentals fields $A_i$ and $B_p$, $i,p=1,2$,  transforming in the representation $(N, \bar N)$ and $(\bar N ,N)$ of the gauge group, respectively,
and the superpotential is 
\be 
W= h\,\epsilon^{ij} \epsilon^{pq} \tr (A_i B_p A_j B_q )\, .
\ee  
The theory has two $SU(2)$ global symmetries transforming the $A_i$ and $B_p$ independently as doublets. There is, in addition, a non-anomalous baryonic symmetry transforming the fields $A_i$ with charge $+1$ and the fields $B_p$ with charge $-1$. 

The classical moduli space is obtained by imposing the conditions (\ref{DFgen}), which in the present case read
\begin{subequations}
\label{DF}
\begin{align}
\epsilon^{pq} B_p A_i B_q & = \epsilon^{ij} A_i B_p  A_j =0\,, \label{DF1}\\
A_1 A_1^\dagger +A_2 A_2^\dagger - B_1^\dagger B_1- B_2^\dagger B_2 & = A_1^\dagger  A_1	 + A_2^\dagger A_2 - B_1 B_1^\dagger - B_2 B_2^\dagger = \calv\,  \bbone\,,   \label{DF2}
\end{align}
\end{subequations}
where the fields are regarded as $N$ by $N$ matrices and
\be\label{calvdef}
\calv\equiv \frac{1}{N}\tr(A_1^\dagger  A_1	 + A_2^\dagger A_2 - B_1 B_1^\dagger - B_2 B_2^\dagger)\,.
\ee
This operator is non chiral but it is contained in the same multiplet of the current that generates the baryonic symmetry.  
Hence its dimension is protected and equal to its classical value, $\Delta_\calv=2$. The expectation value of $\calv$ determines the arbitrary parameter
%\footnote{We recall that the gauge group is  $SU(N)\times SU(N)$ and not $U(N)\times U(N)$, so the D-flatness does not require the vanishing of the trace ${\rm Tr} \sum_i A_i^\dagger A_i - \sum_p B_p B_p^\dagger$.} 
\be\label{conifFI}
\xi=\langle \calv\rangle\,,
\ee
which  is formally equivalent to a FI  for the $U(N)\times U(N)$ theory. As in Section \ref{sec:CFT}, we will refer to $\xi$ as a FI modulus.

Let us first discuss the  mesonic moduli space (\ref{DFgen22}), which is obtained by setting $\xi$=0. This is the subvariety of the  moduli space which can be detected by purely 
mesonic operators 
\be\label{mesop}  {\rm Tr} A_{i_1} B_{p_1} \cdots A_{i_n} B_{p_n} \, ,\ee
with zero baryonic charge.  They are fully symmetric in the indices $i_1,\cdots, i_n$ and $p_1,\cdots,p_n$ by the F-flatness relations (\ref{DF1}). The mesonic operators can be constructed by using as building blocks the four  $N$ by $N$ matrices
\be\label{mesonicbb} \Phi_U = A_1 B_1 \, , \qquad   \Phi_V = A_2 B_2\, , \qquad \Phi_X = A_2 B_1\, , \qquad \Phi_Y= A_1 B_2  \,,
\ee
which transform in the adjoint representation of the first group $U(N)$ and then have zero baryonic charge. Using the F-flatness relations (\ref{DF1}), one can easily check that they 
commute and satisfy the algebraic equation of the conifold as an algebraic variety
\be 
\Phi_U \Phi_V = \Phi_X \Phi_Y \, . 
\ee
Since $\Phi_U, \Phi_V, \Phi_X ,\Phi_Y$ commute, they can be simultaneously diagonalised. The corresponding $N$ eigenvalues take values  in the space defined by the coordinates $(U,V,X,Y)\in\mathbb{C}^4$ satisfying the equation
\be\label{conifold}
UV=XY\,.
\ee
This equation defines the singular conifold. We see that the mesonic moduli space is the symmetric product of $N$ copies of the singular conifold and it has dimension $3N$,
in agreement with (\ref{mesFI}). 

On the other hand, according to equation (\ref{dimmod}), the full moduli space has dimension $3N+1$, which is parametrised not only by the mesonic operators (\ref{mesop}) but also
by baryonic ones. Since the fields transform in the bi-fundamental representation, we can construct gauge-invariant baryonic operators, the prototype being
\be\label{conbaryons2} 
\begin{aligned}
\calb^{A}_{n}&\equiv\frac{1}{N!}\epsilon^{a_1\ldots a_N}\epsilon_{b_1\ldots b_N} (A_1)^{b_1}_{a_1}\cdots (A_1)^{b_{N-n}}_{a_{N-n}}(A_2)^{b_{N-n+1}}_{a_{N-n+1}}\cdots (A_2)^{b_N}_{a_N} \,,\\
\calb^{B}_n&\equiv\frac{1}{N!}\epsilon^{a_1\ldots a_N}\epsilon_{b_1\ldots b_N} (B_1)^{b_1}_{a_1}\cdots (B_1)^{b_{N-n}}_{a_{N-n}}(B_2)^{b_{N-n+1}}_{a_{N-n+1}}\cdots (B_2)^{b_N}_{a_N}  \,,\\
\end{aligned}
\ee
with $n=0,1,\ldots, N$. $\calb^{A}_{n}$ and $\calb^{B}_{n}$ carry baryonic charge $N$ and $-N$, respectively.
We can generalised the above operators, by replacing each single entry in the epsilon contraction with a more general composite field with the 
same transformation properties under the gauge group, for example
\be  (A_{i})^{b_1}_{a_1} \, \qquad \rightarrow \,  \qquad  ( A_{i_1} B_{p_1} \cdots A_{i_k} B_{p_k} A_{i})^{b_1}_{a_1}  \, , \ee
and similarly for $(B_{p})^{b_1}_{a_1}$. This gives a pletora of baryonic operators which are obtained by dressing the elementary baryons (\ref{conbaryons2}) with
mesonic excitations. Mesonic and baryonic operators are not all independent and satisfy many relations. \footnote{Since $\Phi_U,\Phi_V,\Phi_X$ and $\Phi_Y$ in (\ref{mesonicbb}) are $N$ by $N$
matrices, mesons consisting of more than $N$ such building blocks can be written in terms of smaller mesons (see for example \cite{Benvenuti:2006qr,Feng:2007ur} for a general discussion). Moreover, using the tensor identity
$ \epsilon^{a_1\cdots a_N} \epsilon_{b_1\cdots b_N} = N!\delta^{a_1}_{[b1} \cdots  \delta^{a_N}_{b_N]}$,
we can transform particular products of baryons into mesons, for example, schematically
\be \label{barproduct} \calb^{A}_{0}\calb^{B}_{0} \sim \mbox{Tr}\, \Phi_U^N + \cdots \ee
We can only do this because the operator on the left hand side has zero baryonic charge.}  The set of generators of the algebra of chiral operators and the
Hilbert series of the moduli space have been investigated in \cite{Butti:2006au,Forcella:2007wk,Butti:2007jv}.

The baryonic operators   can see directions in the moduli space which are invisible to the mesonic operators,  `resolving' the conifold singularity of mesonic moduli space. 
To have an idea of how this happens, consider the vacua where  the vev of any mesonic operator vanishes.
This requires that either $A_i$ or $B_p$ are zero.
Consider for example the case where all $B_p=0$. The F-flatness conditions (\ref{DF1}) are
automatically satisfied.  The D-flatness conditions (\ref{DF2}) give
\be A_1 A_1^\dagger +A_2 A_2^\dagger   = A_1^\dagger  A_1	 + A_2^\dagger A_2  = \xi\,  \bbone \ee
and necessarily $\xi>0$. We see that, by modding by the gauge transformation, these equations imply that the $N$ eigenvalues of the operators $A_1,A_2$ describe $N$ points moving on a $\mathbb{P}^1$. The $\mathbb{P}^1$ is the exceptional cycle that resolves the conifold singularity. 
Correspondingly, the vevs of the $N+1$ baryonic operators $\calb^A_n$ are generically non-vanishing. These parametrise the  $N$ points moving on the $\mathbb{P}^1$ together with an additional complex modulus which combines  the FI modulus and the Goldstone boson associated with the spontaneously broken baryonic $U(1)$. 
As we will see in the following subsections, all these CFT aspects have a clear holographic counterpart.

\subsection{The dual background}

The generic vacuum of the KW theory is holographically dual to a IIB solution of the kind described in Section \ref{sec:sugravacua}, with the  resolved conifold as internal space $X$. The boundary is then given by the Sasaki-Einstein space $Y=T^{1,1}$.
$X$ has a complex structure which is most easily described  by using toric homogenous coordinates $(Z^1,Z^2,Z^3,Z^4)\in\mathbb{C}^4$, which must be identified under a  $U(1)$ action  with charge vector $Q=(1,1,-1,-1)$ and must satisfy the D-flatness condition
\be\label{confxi}
|Z^1|^2+|Z^2|^2-|Z^3|^3-|Z^4|^2=\xi\,.
\ee
For $\xi=0$ one gets the singular conifold, while there are two possible resolutions associated with  $\xi>0$ or $\xi<0$ respectively, which are related by a flop transition. By comparing this description with the dual CFT, we see that the complex coordinates $(Z^1,Z^2,Z^3,Z^4)$ are naturally associated with the elementary chiral operators $A_1,A_2,B_1,B_2$ in the chosen order. So, as the notation suggest, $\xi$ can be identified with the FI modulus of the dual CFT defined in (\ref{conifFI}). In the following we will assume $\xi > 0$. As a complex space, this resolved conifold $X$ can be represented as
\be\label{confoldcomplex}
X\simeq \frac{\mathbb{C}^4-\{Z^1=Z^2=0\}}{\mathbb{C}^*}\,,
\ee
where $\mathbb{C}^*\simeq U(1)_{\mathbb{C}}$ acts as follows: $(Z^1,Z^2,Z^3,Z^4)\mapsto (\alpha Z^1,\alpha Z^2,\alpha^{-1}Z^3,\alpha^{-1}Z^4)$ for $\alpha\in\mathbb{C}^*$.

The resolved conifold space $X$ has Betti numbers $b_2(X)=b_3(Y)=1$ and $b_4(X)=0$. In particular, $H_2(X;\mathbb{Z})$ is generated by the two-sphere $\mathbb{P}^1$ defined by $Z^3=Z^4=0$. In fact, as a complex space, $X$ can be alternatively identified with the total space of the bundle 
\be\label{resconline}
\calo_{\mathbb{P}^1}(-1)\oplus \calo_{\mathbb{P}^1}(-1)\,.
\ee
 On the other hand, in the above toric description the space $X$ 
inherits also a K\"ahler structure from the ambient flat metric on $\mathbb{C}^4$. This does not coincides with the Ricci-flat  K\"ahler form $J$ on $X$, but lies in the same cohomology class. This allows to compute
\be\label{intJxi}
\int_{\mathbb{P}^1} J=\xi\,,
\ee 
which shows how $\xi$ measures the size of the resolution  $\mathbb{P}^1$. 

We can identify four toric divisors $D_A=\{Z^A=0\}$. Notice that 
$D_1\simeq D_2 \simeq \mathbb{C}^2$ and  $D_3\simeq D_4\simeq \calo_{\mathbb{P}^1}(-1)$. 
In other words $D_1$ and $D_2$ can be identified with the fiber of (\ref{resconline}), while $D_3$ and $D_4$ are obtained by setting to zero one of the two line coordinates in $\calo_{\mathbb{P}^1}(-1)$. Furthermore, these toric divisors define relative homology classes $[D_A]\in H_4(X,Y;\mathbb{Z})$ which are identified as follows $[D_1]=[D_2]=-[D_3]=-[D_4]$. \footnote{As usual, the homological relations between toric divisors can be refined into linear equivalences, see e.g.\ \cite{Cox:2000vi}. For instance, $D_1+D_3$ is represented by the zero-locus of $Z^1Z^3$, which defines a holomorphic function on $X$. Hence $D_1$ and $D_3$ are linearly equivalent, the divisor $D_1+D_3$ corresponds to a trivial line bundle and is then homologically trivial.} 

Let us introduce the harmonic form $\omega$ which is Poincar\'e dual to, say, $D_1$ (or $D_2$). Then, according to our general discussion -- see equation (\ref{Jdec}) -- we can decompose the K\"ahler form on $X$ as follows
\be
J=J_0+v\,\omega\,,
\ee
where $J_0$ is an exact two-form. Since $\int_{\mathbb{P}^1}\omega=\mathbb{P}^1\cdot D_1=1$, from (\ref{intJxi}) we see that we can in fact identify the FI parameter $\xi$ with the (unique) K\"ahler modulus $v$:
\be
v\equiv \xi\,.
\ee
 Hence, in particular, we have
\be\label{caluv}
\langle \calv\rangle =v\,. 
\ee 

The moduli space (\ref{globalM}) for the KW background has the following structure. $\calm_0$ is a $U(1)$ fibration over $\calm_{\rm D3}=\text{Sym}^NX$, with fibral angular variable $\phi$ and local complex coordinates $z^i_I$ along $\calm_{\rm D3}$. The cone $\calk$ coincides with the one-dimensional K\"ahler cone $\mathbb{R}^+$ parametrised by $v$.\footnote{This can be extended to the entire real line by adding the other possible small resolution. In this case, the extended $\calk$ is divided in two chambers, connected by a flop transition.} The coordinates $(v,\phi)$ are symplectically paired and are Legendre dual to a single chiral coordinate $\rho$, as described in general in Section \ref{sec:Mstructure}. Hence, the HEFT will be described by a total of $1+3N$ chiral fields $\rho, z^i_I$. 

In order to compute the HEFT of the KW model we  need the explicit form of $J$ in complex coordinates. This can be described by identifying $X$ with   (\ref{resconline}) and using two local patches $\calu_\pm$ as follows. First introduce two local patches of the base  $\mathbb{P}^1$, parametrised by two local coordinates $\chi$ and $\chi'$, such that $\chi'=1/\chi$, so that $\chi=0$ can be identified with the North pole and $\chi'=0$ with the South pole.  The local patches $\calu_\pm$ on $X$ are then provided by the restriction of the fibration (\ref{resconline}) to these patches on the base $\mathbb{P}^1$. In particular,  $(\chi,U,Y)$ and $(\chi',X,V)$ denote the coordinates on $\calu_+$  and $\calu_-$ respectively, where $(U,Y)$ and $(X,V)$ are fibral coordinates along the vector bundle (\ref{resconline}), related by $X=\chi U$ and $V=\chi Y$.\footnote{\label{foot:localcon} In terms of the homogeneous coordinates, $\calu_+=\{Z^1\neq 0\}$ with $(\chi=\frac{Z^2}{Z^1}, U=Z^1Z^3,Y=Z^1Z^4)$, and $\calu_-=\{Z^2\neq 0\}$ with $\chi'=\frac{Z^1}{Z^2}, X=Z^3Z^2,V=Z^2Z^4\}$.} They satisfy the constrain $XY-UV=0$ and then parametrise the singular conifold. By expressing $(U,V,X,Y)$ in terms of the homogeneous coordinates -- see footnote \ref{foot:localcon} -- it is clear that their values  at position of the $N$ D3-branes correspond to the eigenvalues of the mesonic operators (\ref{mesonicbb}).   

We now introduce  the  radial coordinate 
\be
\begin{aligned}
s=\sqrt{(1+|\chi|^2)(|U|^2+|Y|^2)}=\sqrt{(1+|\chi'|^2)(|X|^2+|V|^2)}\,.
\end{aligned}
\ee
The resolved $\mathbb{P}^1$ then sits at zero radius $s=0$.
The K\"ahler form $J$ is specified  by the (locally defined) K\"ahler potential $k(z,\bar z;v)$, such that $J=\ii\del\delbar k$. In the patch $\calu_+$, it is given by \cite{Candelas:1989js,PandoZayas:2000sq}
\be\label{kconifold}
k(z,\bar z;v)=\frac12\int^{s^2}_0\frac{\d x}{x} G(x;v)+\frac{1}{2\pi} v\,\log(1+|\chi|^2)\,,
\ee
and by replacing $\chi$ with $\chi'$ one gets the K\"ahler potential $k$ on $\calu_-$.
The function $G(x;v)$ is uniquely determined by the equation
\be\label{Gequ}
G(x;v)^3+\frac{3v}{2\pi} \, G(x;v)^2-x^2=0\,,
\ee
and it is explicitly given by \cite{PandoZayas:2000sq}\footnote{
By using the cubic root $(-)^{\frac13}=e^{\frac{\ii\pi}{3}}$, the solution (\ref{defGconif}) remains valid (and real) even  if $x^2<\frac{v^3}{2\pi^3}$.}
\be\label{defGconif}
\begin{aligned}
G(x;v)&=-\frac{1}{2\pi}v +\frac{v^2}{4\pi^2}\caln^{-\frac{1}{3}}(x;v)+\caln^{\frac{1}{3}}(x;v)\\
&\text{with}\quad\caln(x;v)=\frac{1}{2}\Big(x^2-\frac{v^3}{4\pi^3}+x\sqrt{x^2-\frac{v^3}{2\pi^3}}\Big)\,.
\end{aligned}
\ee
For small and large $x/v^{\frac32}$ we have, respectively,
\be\label{expG}
G(x;v)\simeq v\Big[\sqrt{\frac{2\pi}{3}}\, \frac{x}{v^{\frac32}}+\calo\Big(\frac{x^2}{v^3}\Big)\Big]\quad~~~,\quad~~~ G(x;v)\simeq v\Big[\frac{x^{\frac23}}{v}-\frac{1}{2\pi}+\calo\Big(\frac{v}{x^{\frac23}}\Big)\Big]\,.
\ee

The harmonic form $\omega=\ii\del\delbar\kappa$ can be obtained by computing the derivative $\frac{\del J}{\del v}$. In $\calu_+$ the associated potential is given by
\be\label{kappacon}
\kappa(z,\bar z;v)=-\frac{1}{4}\int^{s^2}_0\frac{\d x}{x}\frac{G(x;v)}{\pi G(x;v)+v}+\frac{1}{2\pi}\log(1+|\chi|^2)-\frac{3}{8\pi}\log v\, \, .
\ee
More in detail, by integrating $\omega=\ii\del\delbar\kappa$ one gets $\kappa(z,\bar z;v)$ up to a $v$-dependent piece. This can be fixed by requiring   the boundary condition (\ref{asymptkappa}), which uses the conical radial coordinate $r$, introduced in (\ref{defrcon}) below. This fixes the form (\ref{kappacon}) for $\kappa(z,\bar z;v)$.
% The last term has been added in order to ensure the boundary condition (\ref{asymptkappa}), which can be checked after going to the radial coordinate $r$ which will be  introduced in (\ref{defrcon}). 
On $\calu_-$, $\kappa$ takes the same form (\ref{kappacon}),  up to replacing $\chi$ with $\chi'$.

We can also compute the potential $k_0$ defined as in (\ref{defk0}) and satisfying (\ref{defk0cond}). This is given by
\be
k_0(z,\bar z;v)=\frac34 G(s^2;v)+\frac3{8\pi} v\,.
\ee
We note that $k_0$ is globally defined on $X$, accordingly with our general discussion.

It is also useful to recall how to write the metric and K\"ahler form in real conical coordinates. These are given by five angular coordinates $(\psi,\phi_1,\theta_1,\phi_2,\theta_2)$ defined by
\be
\begin{aligned}
\chi=e^{-\ii\phi_2}\tan\frac{\theta_2}{2}\,,\ U=s\, e^{\frac{\ii}{2}(\psi+\phi_1+\phi_2)}\cos\frac{\theta_1}{2}\cos\frac{\theta_2}{2}\,,\ 
Y=s\, e^{\frac{\ii}{2}(\psi-\phi_1+\phi_2)}\sin\frac{\theta_1}{2}\cos\frac{\theta_2}{2}\,,
\end{aligned}
\ee
and a new radial coordinate $r$ such that
\be\label{defrcon}
r^2=\frac32 G(s^2;v)\,.
\ee
In these coordinates the internal metric in (\ref{10dmetric2}) reads
\be\label{confmetric}
\d s^2_X=t^{-1}(r)\d r^2+ t(r)r^2 \eta^2+\frac16 r^2(\d\theta_1^2+\sin^2\theta_1\d\phi^2_1)      +\Big(\frac16\,r^2+4\pi v \Big)(\d\theta_1^2+\sin^2\theta_1\d\phi^2_1)
\ee
with $t(r)=\frac{4\pi r^2+9v}{4\pi r^2+6v}$ and
$\eta\equiv \frac13\left(\d\psi+\cos\theta_1\d\phi_1+\cos\theta_2\d\phi_2\right)$,
while the K\"ahler form becomes
\be
J=r\d r\wedge \eta+\frac16 r^2\text{vol}^1_{S^2}+\Big(\frac16 r^2+4\pi a^2\Big)\text{vol}^2_{S^2}\,,
\ee
where $\text{vol}^1_{S^2}=\sin\theta_1\d\phi_1\wedge\d\theta_1$ and $\text{vol}^2_{S^2}=\sin\theta_2\d\phi_2\wedge\d\theta_2$. 
The metric (\ref{confmetric}) has conical asymptotically behaviour $\d s^2_X\simeq \d r^2+r^2\d s^2_{T^{1,1}}$, where 
\be
\d s^2_{T^{1,1}}=\eta^2+\frac16 (\d\theta_1^2+\sin^2\theta_1\d\phi^2_1)      +\frac16(\d\theta_1^2+\sin^2\theta_1\d\phi^2_1)
\ee 
is the Sasaki-Einstein metric on $T^{1,1}$, with contact form $\eta$. In conical coordinates the harmonic form $\omega$ takes the form
\be
\omega=-\frac{18 v}{(4\pi r^2+6v)^2}r\d r\wedge\eta-\frac{ r^2}{8\pi r^2+12 v}\text{vol}^1_{S^2}+\frac{4\pi r^2+12 v}{8\pi(4\pi r^2+6v)}\text{vol}^2_{S^2}\,.
\ee
One can  check that $\omega$ satisfy (\ref{asymphatomega2}) and is then $L^{\rm w}_2$-normalisable. 

The KW model has two marginal parameters: the axio-dilaton $\tau$ and the parameter $\lambda$ which sets the (non-dynamical) value of the two-form potentials: $C_2-\tau B_2=\ell^2_{\rm s}\,\lambda \,\omega$.     

\subsection{The HEFT}

We are now ready to derive the HEFT. We have already said that in addition to the $3N$ chiral moduli $z^i_I=(\chi_I,U_I,Y_I)$ (in the patch $\calu_+$) describing the positions of the D3-branes, there is just one  chiral modulus $\rho$.

The implicitly defined K\"ahler potential is given by 
\be\label{EFTKconifold}
K(\rho,\bar\rho, z,\bar z)=2\pi \sum_Ik_0(z_I,\bar z_I;v)=\frac{3\pi}{2}\sum_I G(s^2_I;v)+\frac{3N}{4} v\,,
\ee 
where $s^2_I\equiv (1+|\chi_I|^2)(|U_I|^2+|Y_I|^2)$.
On the r.h.s.\ of (\ref{EFTKconifold}) $v$ must be considered as the function of $(\rho,\bar\rho,z_I,\bar z_I)$ that is obtained by inverting
\be\label{rhoconif}
\Re\rho= -\frac{1}{8}\sum_I\int^{s_I^2}_0\frac{\d x}{x}\frac{G(x;v)}{\pi G(x;v)+v}+\frac{1}{4\pi}\sum_I\log(1+|\chi_I|^2)-\frac{3N}{16\pi}\log v\,,
\ee 
see (\ref{rhoa0}).  From the K\"ahler potential (\ref{EFTKconifold}) one can then derive  the HEFT non-linear sigma model (\ref{rigidkineticaxion0}):
\be\label{kineticC3Z3}
\call_{\rm chiral}=-\frac{\pi}{ \calg(\rho,\bar\rho,z,\bar z)}\nabla\rho\wedge*\nabla\bar\rho- 2\pi\sum_I g_{i\bar\jmath}(z_I,\bar z_I;v)\d z_I^i\wedge*\d \bar z_I^{\bar\jmath}+\text{(fermions)}\,,
\ee
where
\be
\calg(\rho,\bar\rho,z,\bar z)=\frac{3}{16\pi}\sum_I\frac{1}{v+\pi G(s^2_I;v)}
\ee
 is obtained from (\ref{derrhov}), and $g_{i\bar\jmath}(z,\bar z;v)=\del_i\delbar_{\bar\jmath} k(z,\bar z;v)$ are the components of the K\"ahler metric (\ref{confmetric}) in complex coordinates. Furthermore, $\nabla_\mu\rho=\del_\mu\rho-\sum_I\cala_i(z_I,\bar z_I;v)\del_\mu z_I^i$ with (see (\ref{defcala}))
\be
\hskip -0.2truecm \cala_i(z,\bar z;v)\d z^i=\frac{1}{4 v+4\pi G(s^2;v)}\left[\frac{2v +\pi G(s^2;v)}{\pi(1+|\chi|^2)}\,\bar\chi\d \chi-\frac{G(s^2;v)(\bar U\d U+\bar Y\d Y)}{|U|^2+|Y|^2}\right] .
\ee

 It is interesting to observe that, as far as the D3-branes are all away from the blown-up $\mathbb{P}^1$ (i.e.\ $s_I^2\neq 0$ for all $I$), the Lagrangian (\ref{kineticC3Z3}) remains regular in the limit $v\ll 1$, in which the internal space $X$ develops a conifold singular. This is true not only for the $z^i_I$ kinetic terms but, maybe unexpectedly, also for the $\rho$ kinetic term. Indeed, this limit is practically implemented by considering $s_I^2\gg v^{\frac32}$ and by using the second of (\ref{expG}) we see that $\calg\simeq \frac{3}{16\pi^2}\sum_I s^{-\frac{4}{3}}_I$, which is  finite. Roughly, the singularity is invisible to the $N$ D3-branes and then the HEFT remains regular even in this limit.

As already remarked, in all above expressions one should consider $v$ as a function of the chiral fields $\rho$ and $z^i_I$. We do not know a general analytic formula for such function, but one can in principle derive it in a perturbative expansion. We can for instance consider  the region in the moduli space in which $v$ is quite large  while $s^2_I$ are finite, so that $s^2_I/v^{\frac32}\ll 1$. Dually, this roughly means that the vevs of the mesonic operators are very small compared to the vev of the operator (\ref{calvdef}). In terms of the HEFT chiral fields, this regime corresponds to  $s^2_I e^{\frac{8\pi\Re\rho}{N}}\ll 1$. By using the first of (\ref{expG}) in  (\ref{rhoconif}), we find that
\be
v=\prod_I(1+|\chi_I|^2)^{\frac{4}{3N}}e^{-\frac{16\pi\Re\rho}{3N}}-\frac1N \Big(\frac{2\pi}{3}\Big)^{\frac32}\prod_I
(1+|\chi_I|^2)^{-\frac{2}{3N}}e^{\frac{8\pi\Re\rho}{3N}}\sum_J s_J^2+\ldots\,,
\ee
where we have neglected terms of order $\sim s^4_I e^{\frac{32\pi\Re\rho}{3N}}$. To this order, the HEFT K\"ahler potential (\ref{EFTKconifold}) takes the following explicit form
\be
K(\rho,\bar\rho,z,\bar z)=\frac{3N}{4}\prod_I(1+|\chi_I|^2)^{\frac{4}{3N}}e^{-\frac{16\pi\Re\rho}{3N}}+\frac32 \Big(\frac{2\pi}{3}\Big)^{\frac32}\prod_I
(1+|\chi_I|^2)^{-\frac{2}{3N}}e^{\frac{8\pi\Re\rho}{3N}}\sum_J s_J^2+\ldots
\ee
and, for instance, the first kinetic prefactor in (\ref{kineticC3Z3}) is explicitly given by 
\be
\frac{\pi}{\calg(\rho,\bar\rho,z,\bar z)}=\frac{16\pi^2}{3N}\prod_I(1+|\chi_I|^2)^{\frac{4}{3N}}e^{-\frac{16\pi\Re\rho}{3N}}+\frac{8\pi^2}{3N^2}\Big(\frac{2\pi}{3}\Big)^{\frac32}\prod_I
(1+|\chi_I|^2)^{-\frac{2}{3N}}e^{\frac{8\pi\Re\rho}{3N}}\sum_J s_J^2+\ldots
\ee

Coming back to the complete HEFT, in addition to being manifestly $\caln=1$ supersymmetric, it should also be invariant under a non-linear realisation of the  superconformal generators that are spontaneously broken by the vacua at which the HEFT is defined. Let us explicitly check it for the dilations.

In order to do that, we must identify the scaling dimensions of the fields entering the HEFT. This is particularly easy in the KW model since, as discussed above,  the CFT chiral fields $A_1,A_2,B_1,B_2$ are in natural correspondence with the homogeneous coordinates $Z^A$. The dimension of the fields $A_1,A_2,B_1,B_2$ in the CFT is uniquely fixed to be $3/4$ by the $SU(2)\times SU(2)$ symmetry and the fact that the quartic superpotential must have dimension $3$. Hence we can assign  to $Z^A$ a scaling dimension $\Delta_{Z}=\frac34$ and this in turn implies the scaling dimensions $\Delta_{\chi_I}=\Delta_{\chi'_I}=0$ and $\Delta_{U_I}=\Delta_{V_I}=\Delta_{X_I}=\Delta_{Y_I}=\frac32$ (see footnote \ref{foot:localcon}), and then also $\Delta_{s^2_I}=3$. The scaling dimension of $\rho$ can be determined by relating it to the the expectation value of the baryonic operators, as we will see in the next subsection.  The result will imply that  $e^{-2\pi\rho}$ has scaling dimension $\frac{3N}{4}$. Furthermore,  the real K\"ahler modulus $v$ has scaling dimension $\Delta_v=2$, as one can immediately conclude from  (\ref{caluv}).\footnote{By using (\ref{rhoconif}) one can check that this is consistent with the scaling dimension $\frac{3N}{4}$ of  $e^{-2\pi\rho}$. }

It is now easy to explicitly check that the K\"ahler potential  (\ref{EFTKconifold}) has scaling dimension $2$, which implies that the supersymmetric Lagrangian $\int\d^4\theta K$ is indeed scale invariant, as required.

\subsection{Baryonic condensates}
\label{sec:KWbaryons}

So far, we have only partially provided a CFT interpretation of the supergravity chiral fields $\rho,z^i_I$. In particular, we have identified the HEFT chiral fields $U_I, V_I, X_I,Y_I$  with the $N$ eigenvalues of the mesonic operators (\ref{mesonicbb}). On the other hand, the expectation value of a general mesonic operator cannot `see' neither $\rho$ nor the position of D3-branes sitting at different points of the resolution $\mathbb{P}^1$. 

This additional information is in fact encoded in the vev of the baryonic operators which, according to the prescription \cite{Klebanov:2007us} reviewed in Section \ref{sec:baryonvev}, can be computed by evaluating the on-shell action of E3-branes on non-compact divisors in $X$. More precisely, the different baryons with given dimension and charge are associated with different states in the Hilbert space which is obtained by quantising the moduli space of the associated divisors, as in \cite{Beasley:2002xv}.   

In particular, the $N+1$ baryons $\calb^A_n$ defined in (\ref{conbaryons2}) naturally correspond to the family of divisors obtained by rotating $D_1$ (or $D_2$). The generic divisor in this family is described by the zero-locus of the polynomial of degree-one:
\be
P^A_{c_1,c_2}(Z^1,Z^2)\equiv c_1 Z^1+c_2 Z^2=0\,.
\ee
The divisor does not change if we rescale $c_1$ and $c_2$ by the same complex number. Hence different divisors 
in this family are parametrised by the point  $[c_1:c_2]$ in a complex projective space $\mathbb{P}^1_{A}$. We then denote this class of divisors by $D^A_{[c_1:c_2]}$. In this notation $D_1\equiv D^A_{[1:0]}$ and $D_2\equiv D^A_{[0:1]}$. Correspondingly, by using the description (\ref{confoldcomplex}) of $X$, the polynomials  $P^A_{c_1,c_2}(Z^1,Z^2)$ are associated with global sections $\zeta^A_{[c_1:c_2]}(z)$ of a non-trivial line bundle over $X$, such that $D_{[c_1:c_2]}=\{\zeta^A_{[c_1:c_2]}(z)=0\}$. 

 In order to evaluate the vev  of the  baryons $\calb^A_{i_1\ldots i_N}$ in terms of the HEFT chiral fields, we  use the semiclassical results of Section  \ref{sec:baryonvev}. All the divisors $D^A_{[c_1:c_2]}$ are diffeomorphic to $\mathbb{C}^2$ and have boundaries $\Sigma^A_{[c_1:c_2]}\simeq S^3$. Hence $b_2(D^A_{[c_1:c_2]})=0$,  $c_1(D^A_{[c_1:c_2]})=0$ and  $H_1(\Sigma^A_{[c_1:c_2]};\mathbb{Z})=0$, so that the corresponding E3-brane can support just a trivial flat connection. Since we have chosen the harmonic form $\omega$ to be Poincar\'e dual to $D_1$, and then to any $D^A_{[c_1,c_2]}$, the general formula (\ref{expVEV}) boils down to 
\be\label{Abaryon}
\langle \calb^A_{[c_1:c_2]}\rangle=\prod_I\zeta_{[c_1:c_2]}(z_I)\,e^{-2\pi\rho}\,,
\ee
On the other hand $\prod_I\zeta_{[c_1:c_2]}(z_I)$ is associated to the homegeneous  polynomial  
\be\label{expoly}
\begin{aligned}
\prod_IP^A_{c_1,c_2}(Z^1_I,Z^2_I)=&\, \sum^N_{n=0}P^A_n \psi_n(c_1,c_2)\,,
\end{aligned}
\ee
where
\be
\psi_n(c_1,c_2) = \frac{N!}{(N-n)!n!}\,  c^{N-n}_{1}c^n_2
\ee
and $P^A_n$ are the polynomials which can be  obtained by inserting the matrices
\be
A_1=\left(\begin{array}{cccc}
 Z^1_1 & 0 & \ldots &       \\
        0   & Z^1_2 & & \\
     \vdots & & \ddots & \vdots\\
     &&\ldots& Z^1_N
              \end{array}\right)\quad ,\quad A_2=\left(\begin{array}{cccc}
 Z^2_1 & 0 & \ldots &       \\
        0   & Z^2_2 & & \\
     \vdots & & \ddots & \vdots\\
     &&\ldots& Z^2_N
              \end{array}\right)
\ee
in $\calb^A_n$ defined in (\ref{conbaryons2}). Correspondingly, in (\ref{Abaryon}) we can expand 
\be\label{exphols}
\prod_I\zeta_{[c_1:c_2]}(z_I)=\sum^N_{n=0}\zeta^A_n(z_1,\ldots, z_N) \psi_n(c_1,c_2)\,,
\ee
where $\zeta^A_n(z_1,\ldots, z_N)$ are the holomorphic sections over $\calm_{\rm D3}={\rm Sym}^NX$, which correspond to the homogeneous polynomials $P^A_n$. For instance, 
\be
\zeta^A_1(z_1,\ldots, z_N)=\left\{ \begin{array}{ll} \frac1N\,(\chi_1 +\chi_2+\ldots +\chi_N)  &\text{in $\calu_+$}\\
\frac1N\,(\chi'_2\cdots \chi'_N +\chi'_1\chi'_3\cdots\chi'_N +\ldots +\chi'_1\cdots\chi'_{N-1})    &\text{in $\calu_-$} \end{array}\right.\,.
\ee

From (\ref{exphols}) we see that $\langle \calb^A_{[c_1:c_2]}\rangle$ is associated with a homogeneous polynomial of degree $N$ in $(c_1,c_2)$. In other words, we can regard $\langle \calb^A_{[c_1:c_2]}\rangle$ as defining a holomorphic wave function taking values in the holomorphic line bundle $\calo_{\mathbb{P}^1_A}(N)$ over $\mathbb{P}^1_A$. But the space of holomorphic sections of $\calo_{\mathbb{P}^1_A}(N)$  exactly corresponds  to the quantum Hilbert space generated by the  baryons $\calb^A_n$ \cite{Beasley:2002xv}. In particular,  these operators are associated with the $N+1$-dimensional basis defined by $\psi_n(c_1,c_2)$. Hence, we can read their expectation values from  (\ref{Abaryon}) by picking the appropriate coefficient in the expansion  (\ref{exphols}). We then arrive at
\be
\langle \calb^A_n\rangle=\zeta^A_n(z_1,\ldots, z_N)\,e^{-2\pi\rho}\,.
\ee
 Note that the above definition of the polynomials $P^A_n$ makes it clear the direct  connection between this quantisation procedure and the dual baryonic operators $\calb^A_n$ defined in (\ref{conbaryons2}).

The computation of the vevs of the operators  $\calb^B_n$ is slightly less straightforward.  The associated family of divisors $D^B_{[c_3:c_4]}$, with $[c_3:c_4]\in\mathbb{P}^1_B$, is now defined by the vanishing of polynomials of degree one
\be
P_{c_3,c_4}(Z^3,Z^4)\equiv c_3 Z^3+c_4 Z^4\,.
\ee
which descend to corresponding holomorphic sections $\zeta^B_{[c_3:c_4]}(z)$ on $X$.
The divisors $D^B_{[c_3:c_4]}$ contain the resolved $\mathbb{P}^1$ and are isomorphic to the total space of the line bundle $\calo_{\mathbb{P}^1}(-1)$. Let us for the moment omit the subscript $_{[c_3:c_4]}$ to simplify the notation.  The divisors $D^B$ have boundary three-cycles $\Sigma^B\simeq S^3$. Then $b_2(D^B)=1$, $H_1(\Sigma^B;\mathbb{Z})=0$ and $H_2(D_B,\Sigma^B;\mathbb{Z})=\mathbb{Z}$, which is generated by a non-compact holomorphic curve $\tilde C$.\footnote{For instance, in  $D^B_{[1:0]}\equiv D_3$, we can take $\tilde C=\{Z_1=Z_3=0\}$.} Its Poincar\'e dual $[\tilde C]\in H^2(D^B;\mathbb{Z})$ is  cohomologous to $\omega|_{D_B}$ and we denote by $\alpha$ its primitive (1,1) harmonic $L_2$-normalisable representative.  

On the other hand, the resolved $\mathbb{P}^1$ generates $H_2(D^B;\mathbb{Z})=\mathbb{Z}$ and is such that $\int_{\mathbb{P}^1}\alpha=\mathbb{P}^1\cdot\tilde C=1$. Since $H_1(\Sigma^B;\mathbb{Z})=0$,  the boundary of $\tilde C$ can be (non-holomorphically) collapsed, getting a(n anti-holomorphic) compact two-cycle which is homologous to $-\mathbb{P}^1$. By regarding $D^B$ as a toric variety, one can compute the first Chern class $c_1(D^B)=[\tilde C]$ and the Euler characteristic $\chi(D^B)=2\mathbb{P}^1\cdot \tilde C=2$. Reintroducing the subscript $_{[c_3:c_4]}$, we arrive at the semiclassical formula 
\be\label{semB}
\begin{aligned}
\langle \calb^B_{[c_3:c_4]} \rangle
=\prod_I\zeta^B_{[c_3:c_4]}(z_I) \cala(\lambda,\tau)\, e^{2\pi\rho}\,\,,
\end{aligned}
\ee
where
\be
\cala(\lambda,\tau)=e^{\frac{\pi\ii}{\Im\tau}\lambda\Im\lambda-\frac{\pi\ii}{6}\tau}\,\Theta\left[\begin{array}{c}  \frac12 \\ 0\end{array}\right](\lambda| \tau)\,.
\ee

We can now quantise the family of divisors $D^B_{[c_3:c_4]}$, as we did for $D^A_{[c_1:c_2]}$. In this way we  extract from (\ref{semB}) the following expectation values of the baryons $\calb^B_n$:
\be
\langle\calb^B_n\rangle=\zeta^B_n(z_1,\ldots,z_N)\,\cala(\lambda,\tau)\, e^{2\pi\rho}\,.
\ee
Here $\zeta^B_n(z_1,\ldots,z_N)$ are holomorphic sections on $\calm_{\rm D3}={\rm Sym}^NX$ 
which correspond to the homogenous polynomials $P^B_n$  obtained by inserting the matrices
\be
B_1=\left(\begin{array}{cccc}
 Z^3_1 & 0 & \ldots &       \\
        0   & Z^3_2 & & \\
     \vdots & & \ddots & \vdots\\
     &&\ldots& Z^3_N
              \end{array}\right)\quad ,\quad B_2=\left(\begin{array}{cccc}
 Z^4_1 & 0 & \ldots &       \\
        0   & Z^4_2 & & \\
     \vdots & & \ddots & \vdots\\
     &&\ldots& Z^4_N
              \end{array}\right)
\ee
in  $\calb^B_n$ defined in (\ref{conbaryons2}). Again we see that, through the quantisation of the divisor moduli space, the precise connection with the dual baryonic operators naturally emerges.

As a simple check, let us move all $D3$ branes on the resolved $\mathbb{P}^1$, defined by $Z^3=Z^4=0$,
so that only $N$ of the $3N$ chiral fields  $z^i_I$ are free to vary.
In this case $\langle \calb^B_n\rangle =0$ for any $n=0,\ldots,N$, while the $N+1$ vevs of $ \langle \calb^A_n\rangle $ are generically non-vanishing. These are in  correspondence with the $N+1$ non-vanishing chiral fields given by $\rho$ and the positions of the $N$ D3-branes on   $\mathbb{P}^1$. As a further particular subcase, suppose that all D3-brane sit at north pole of the $\mathbb{P}^1$, defined by $Z^1=Z^3=Z^4=0$. In this case only $\calb^A_N$ is non-vanishing and, by using (\ref{rhoconif}),  $|\calb^A_N|\simeq v^{\frac{3N}{8}}$, reproducing the result of  \cite{Klebanov:2007us}.

%%%%%%%%%%%%%%%%%%%%%%%%%%%%%%%%%%%%%%%%%%%%%%%%%%%%%%%%%%%%%%%%%%%%%%%%%%%%%%%%%%%%%%%%%%%%%%%%%%%%
%%%%%%%%%%%%%%%%%%%%%%%%%%%%%%%%%%%%%%%%%%%%%%%%%%%%%%%%%%%%%%%%%%%%%%%%%%%%%%%%%%%%%%%%%%%%%%%%%%%%

\section{Discussion}
\label{sec:conclusions}

In this paper we have identified the holographic effective field theory (HEFT) describing the low-energy dynamics of a large class of strongly-coupled $\caln=1$ CFTs at the generic points of their moduli space, at which the superconformal symmetry is spontaneously broken. These CFTs corresponds to IR fixed points of quiver gauge theories engineered by placing  $N$ D3-branes at the tip of a Calabi-Yau cone $C(Y)$ over a Sasaki-Einstein space $Y$. Our HEFT is  defined, at the two-derivative order, in terms of a non-trivial K\"ahler potential for an appropriate set of chiral fields, which parametrise  the open and closed string moduli of the dual holographic background. We have outlined how these HEFT chiral fields determine the vev of the CFT chiral fields. In particular, we have provided a semiclassical formula for the vev of  baryonic operators, extending the results of \cite{Klebanov:2007us,Martelli:2008cm}. We have also provided an alternative description of the geometry of the moduli space determined by the HEFT, in terms of a mixed complex-symplectic potential, whose geometrical interpretation is more transparent and which is more directly connected with the classical description of the CFT moduli space. Our general results have been explicitly applied to the Klebanov-Witten model \cite{Klebanov:1998hh}.

In the paper we have mostly assumed to be at the generic point in the moduli space, at which the D3-branes are separated. 
On the other hand, our HEFT breaks down once some D3-branes coincide. Indeed, we know that at these points the low-energy theory must contain some $\caln=4$ SYM sectors.
For instance, suppose that all $N$ D3-branes coincide at a point of coordinates $z^i_{\rm c}$. The supergravity background is  well defined and close to the coinciding D3-brane it develops a mildly curved AdS$_5\times S^5$ background supporting $N$ units of $F_5$ flux, as in \cite{Klebanov:2007us}.  Such throat corresponds to the appearance of a $SU(N)$ $\caln=4$ SYM theory in the IR, to which the UV CFT flows. On the other hand, as it is clear from the holographic description, the closed string moduli and the open string center-of-mass moduli  $z^i_{\rm c}$ should still appear as dynamical degrees of freedom in a low-energy effective theory. Assuming that the dynamics of the $\caln=4$ SYM sector decouples from the moduli dynamics, one may derive an HEFT for the latter just by substituting all $z_I^i$ with $z^i_{\rm c}$ in the formulas obtained in the present paper. Clearly, this procedure can be adapted to more general cases in which the D3-branes form smaller groups.

The HEFT has been derived starting from the ten-dimensional supergravity and performing a tree-level dimensional reduction.
Hence, a priori, it is valid only for small string coupling and small curvatures. While we can always justify the tree-level approximation by choosing a small enough string coupling $g_{\rm s}$, which is a non-dynamical marginal parameter,  the curvature corrections may become important in some region of the moduli space. In particular, the internal space of the string background is provided by a smooth Calabi-Yau resolution $X$ of $C(Y)$, hosting $N$ back-reacting D3-branes. There are then two sources of curvature: one  associated with curvature of the underlying Calabi-Yau metric itself and one associated with the warping produced by the D3-branes. 

Let us first focus on the Calabi-Yau curvature. The K\"ahler moduli $v^a$ measure,  in string units,  the Einstein-frame volumes of the two-cycles present in the smooth space $X$. Then, the ten-dimensional supergravity approximation we started from is expected to be valid only when the corresponding string-frame volumes are large in string units, i.e.\ $  v^a\gg 1/\sqrt{g_{\rm s}}$. On the other hand, at the level of the HEFT such condition is not sensible, because of the underlying conformal symmetry. 
Indeed, if all $v^a$ are non-vanishing, by an appropriate dilation we can always make them arbitrarily large.  Since under this dilation the complete HEFT must be invariant, we can always choose to compute it in the regime in which all   $  v^a$ are  large enough   and the Calabi-Yau geometry is well described by the leading ten-dimensional supergravity.

We can now turn to the warping.  Asymptotically, the warping produces the AdS$_5\times Y$ geometry with string frame radius $R_{\rm st}\sim \ell_{\rm s}(g_s N)^\frac14$. As usual, the conditions  $g_{\rm s}\ll 1$ and  $R_{\rm st}/ \ell_{\rm s}\gg 1$ require the large-$N$ limit with large `t Hooft coupling $\lambda_{\rm YM}=g_{\rm s}N$, which may be interpreted as a diagonal combination of the quiver gauge couplings.  As one moves closer to the  D3-branes, in the generic vacua at which they are not coincident, the space develops $N$ local strongly curved  AdS$_5\times S^5$ throats. Even if AdS$_5\times S^5$ is an exact string background \cite{Berkovits:2004xu}, one may wonder whether higher order corrections due to such strongly curved warping can affect the HEFT. We do not have a definitive answer to this question. However, we observe that the warping enters as an `integrated' quantity in  the HEFT,  effectively  disappearing from it and leaving just the dependence on the  positions of the D3-branes which source it. Hence, our HEFT does not `see' such localised divergences. 

To further support this idea, we observe that the dilation discussed above stretches also the 
distance between the  D3-branes. This means that, generically, we can assume that the Calabi-Yau radius of curvature and the mutual distance between the non-coincident D3-branes is much larger than the string length $\ell_{\rm s}$. In this case, since the strongly curved regions are localised around the D3-branes, each  D3-brane should be well approximated by a probe D3-brane on a weakly curved background generated by the remaining $N-1$ D3-branes. By consistency, our  HEFT should then reproduce the kinetic metric for the moduli $z^i_I$ obtained by considering the $I$-th D3-brane as a probe. Indeed, by expanding the corresponding DBI action one gets  $-2\pi g_{i\bar\jmath}(z_I,\bar z_I)\del_\mu z_I^i\del^\mu \bar z_I^{\bar\jmath}$. Notice that  any explicit dependence on the warping has dropped out and so the probe D3-brane `sees' only the underlying Calabi-Yau metric. 
This happens  basically because of the mutual BPS-ness of the D3-branes. % and so we expect such kinetic term to be quite robust. 
%But exactly the same kinetic term  appears in  second term of 
We see that HEFT Lagrangian (\ref{rigidkineticaxion0})  perfectly matches the probe expectation.  

The above observations suggest that our second derivative HEFT may in fact admit a broader regime of validity than naively expected. 
%and it may be even be exact in the strict large-$N$ limit.
 It would be very interesting to check this possibility more explicitly, by directly studying the implication of the perturbative higher derivative contributions to the ten-dimensional supergravity. Another source of correction could come from non-perturbative corrections arising from various kinds of world-sheet or brane instantons. In this respect, it would be important to inspect in detail other explicit models, which for instance include anomalous baryonic symmetries. Indeed, in such cases $b_4(X)\neq 0$ and there could be potential corrections arising from supersymmetric D3-brane instantons.   

 Furthermore, our approach implicitly assumes that, at sufficiently low energies, our HEFT massless fields are decoupled from the  massive four-dimensional states which would be associated to possible normalisable non-zero modes of the internal supergravity configuration. It would be interesting to  investigate the spectrum of such non-zero modes and more explicitly study their impact on the HEFT.

Finally, we observe that the methods of the present paper can have a broader range of potential applications. For instance, they have an obvious counterpart for the holographic models which are dual to $\caln=2$ three-dimensional CFTs. Furthermore, the holographic string backgrounds can be considered as local strongly warped regions of proper compactifications and indeed our HEFTs can be generalised to describe  local sectors of phenomenologically motivated string models.

%%%%%%%%%%%%%%%%%%%%%%%%%%%%%%%%%%%%%%%%%%%%%%%%%%%%%%%%%%%%%%%%%%%%%%%%%%%%%%%%%%%%%%%%%%%%%%%%%%%%
%%%%%%%%%%%%%%%%%%%%%%%%%%%%%%%%%%%%%%%%%%%%%%%%%%%%%%%%%%%%%%%%%%%%%%%%%%%%%%%%%%%%%%%%%%%%%%%%%%%%

\vspace{1cm}

\centerline{\bf \large Acknowledgements}

\bigskip

\noindent The authors are grateful to M.~Bianchi, R.~Minasian, F.~Morales and L.~Vecchi  for discussions. The work of L.M.~is partially supported by the Padua University Project CPDA144437. A.Z.~is supported by the INFN and the MIUR-FIRB grant RBFR10QS5J ``String Theory and Fundamental Interactions''.

\vspace{2cm}

\centerline{\LARGE \bf Appendices}

\begin{appendix}

%%%%%%%%%%%%%%%%%%%%%%%%%%%%%%%%%%%%%%%%%%%%%%%%%%%%%%%%%%%%%%%%%%%%%%%%%%%%%%%%%%%%%%%%%%%%%%%%%%%%
%%%%%%%%%%%%%%%%%%%%%%%%%%%%%%%%%%%%%%%%%%%%%%%%%%%%%%%%%%%%%%%%%%%%%%%%%%%%%%%%%%%%%%%%%%%%%%%%%%%%

\section{HEFT from $M_{\rm P}\rightarrow\infty$ limit}
\label{app:rigidlimit}

In this appendix we derive the effective Lagrangian (\ref{rigidkineticaxion0}) by taking the rigid limit of the effective field theory of warped compactifications derived in \cite{Martucci:2014ska}. 
The following discussion can be applied to  quite general local models, not necessarily restricted to the class considered in this paper.

\subsection{Warped EFT for finite $M_{\rm P}$}
\label{app:finiteMP}

We first summarise some key points of  \cite{Martucci:2014ska}, which focuses on the  IIB/F-theory  warped flux compactifications  discussed  in \cite{GKP}.
 The Einstein frame metric has the form
\be\label{10dmetric}
\ell^{-2}_{\rm s}\d s^2_{10}=e^{2A}|\Phi|^2\d s^2_{\mathbb{M}^{1,3}}+e^{-2A}\,\d s^2_{X}\,,
\ee
where  $\d s^2_{X}=g_{i\bar\jmath}\,\d z^i \d \bar z^{\bar\jmath}$ is a K\"ahler metric over the internal space $X$, which is {\em compact}, and $\Phi$ plays the role of conformal compensator.
The metric $\d s^2_{X}$ is normalised to give a fixed finite volume
\be\label{const}
{\rm v}_0=\int_X \d\text{vol}_X=\frac1{3!}\int_{X} J\wedge J\wedge  J\,,
\ee
where 
\be
J=\ii g_{i\bar\jmath}\,\d z^i\wedge \d \bar z^{\bar\jmath}\,,
\ee
 is the associated K\"ahler form.
 The warp factor  must satisfy the Poisson-like equation
\be\label{poisson}
\Delta e^{-4A}=\frac{1}{\ell^4_{\rm s}}* Q_6\,,
\ee
where
\be
Q_6=\ell^4_{\rm s}\sum_{I\in\text{D3's}}\delta^6_I+Q^{\rm bg}_6\,,
\ee
with 
\be
Q^{\rm bg}_6=F_3\wedge H_3-\frac14\ell^4_{\rm s}\sum_{O\in \text{O3}'s}\delta^6_O+\ldots
\ee
containing additional sources for the warping. The tadpole conditions requires  no net D3-brane charge: $\int_X Q_6=0$. 
The general solution of (\ref{poisson}) can be written as
\be\label{warpingsplit}
e^{-4A}=a+e^{-4A_0}\,,
\ee
where $a$ is an arbitrary constant, the ``universal modulus", and $e^{-4A_0}$ is the particular solution of (\ref{poisson}) 
such that \footnote{The notation may be misleading, since the function $e^{-4A_0}$ can become negative in some regions of the internal space. } 
\be\label{normcond0}
a=\frac{1}{{\rm v}_0}\int_X e^{-4A}\d\text{vol}_X\,.
\ee

In addition to the universal modulus $a$, there are other $h^{1,1}-1$   K\"ahler moduli, which are identified by expanding the K\"ahler form in a basis of integral 
harmonic $(1,1)$ forms $\omega_A\in H^2(X;\mathbb{Z})$:
\be
J=v^A\omega_A\,.
\ee
They are constrained by the condition (\ref{const}), which can be rewritten as
\be\label{const1}
\frac1{3!}\,\cali_{ABC}v^Av^Bv^C={\rm v}_0\,,
\ee
where $\cali_{ABC}\equiv \int_X\omega_A\wedge \omega_B\wedge\omega_C$ are triple intersection numbers.

There are also $3N$ complex moduli $z^i_I$, $I=1,\ldots,N$, parametrising the position of $N$ mobile D3-branes in the internal space. 
 For the purposes of the present paper, we can consider the axio-dilaton and complex structure moduli as frozen, while there may be additional axionic moduli, associated with the $C_2$, $B_2$ and the seven-brane Wilson lines. We will be interested only in the $C_2$ and $B_2$ moduli. However, in order to simplify the presentation, we initially assume that they are absent.

As explained in \cite{Martucci:2014ska}, the K\"ahler potential is just given by
\be\label{compactK}
K=-3\log (4\pi {\rm v}_0 a)\,.
\ee
The definition of the proper chiral fields $\rho_A$ parametrising the K\"ahler deformations (and the axionic partners) requires the introduction of
a set of  (locally defined) `potentials'  $\kappa_A(z,\bar z;v)$ such that 
\be
\omega_A=\ii\del\delbar\kappa_A\,.
\ee
In order to derive the effective action D-terms arising from (\ref{compactK}), one only needs the explicit form of the real part of the chiral fields $\rho_A$, which is given by
\be\label{defchiral}
\Re\rho_A=\frac12 a\,\cali_{ABC}v^Bv^C+\frac12 \sum_I \kappa_A(z_I,\bar z_I;v)+h_A(v)\,,
\ee
with 
\be
h_A(v)\equiv \frac{1}{2\pi\ell^4_{\rm s}}\int_X(\pi\kappa_A-\Re\log\zeta_A)Q^{\rm bg}_6\,,
\ee
where $\zeta_A(z)$ is a holomorphic section of the holomorphic line bundle whose first Chern class equals $\omega_A$.

One can then show that the bosonic four-dimensional Lagrangian computed from the K\"ahler potential (\ref{compactK}) is
\be\label{kinetic}
\call=-\frac{1}{4{\rm v_0}a}M^2_{\rm P}\calg^{AB}\nabla\rho_A\wedge *\nabla\bar\rho_B-\frac{1}{2{\rm v_0}a}M^2_{\rm P}\sum_I g_{i\bar\jmath}(z_I,\bar z_I)\d z_I^i\wedge *\d \bar z_I^{\bar\jmath}
\ee
where 
\begin{subequations}
\begin{align}
\calg^{AB}&\equiv\frac{1}{2{\rm v}_0 a}v^Av^B-(M_{\rm w}^{-1})^{AB}\label{calg}\,,\\
\nabla\rho_A&\equiv\d\rho_A-\sum_I\cala^{I}_{Ai}\d z^i_I\,,\\
\cala^{I}_{Ai}&\equiv\frac{\del \kappa_A(z_I,\bar z_I;v)}{\del z^i_I}\,.
\end{align}
\end{subequations}
Here $(M_{\rm w}^{-1})^{AB}$ is the inverse of 
\be
M_{{\rm w}AB}=\int_Xe^{-4A}J\wedge \omega_A\wedge\omega_B\,,
\ee
and the four-dimensional Planck mass $M_{\rm P}$ is related to the ten-dimensional metric (\ref{10dmetric}) by the formula 
\be\label{MP}
M^2_{\rm P}=4\pi {\rm v}_0 a|\Phi|^2\,.
\ee

%%%%%%%%%%%%%%%%%%%%%%%%%%%%%%%%%%%%%%%%%%%%%%%%%%%%%%%%%%%%%%%%%%

\subsection{Dual formulation with linear multiplets}

Eventually, we want to take the decompactification/$M_{\rm P}\rightarrow\infty$ limit of the flux compactifications described in subsection \ref{app:finiteMP}. As we will see, such limit is more naturally described in the dual formulation in terms of linear multiplets $(l^A,H^A)$, with $l^A$ real scalars and $H^A=\d b^A$ real 3-forms, which are dual to  the chiral multiplets $\rho_A$.  The scalar component $l^A$ is related to $\Re\rho_A$ by (see for instance \cite{Grimm:2005fa} for a review) 
\be\label{linearl}
l^A=-\frac14\frac{\del K}{\del\Re\rho_A}=-\frac{v^A}{4{\rm v}_0 a }\,,
\ee
which shows that $l^A$ has a simple geometrical interpretation. In terms of the linear multiplets, the effective bosonic Lagrangian becomes  
\be\label{dualkinetic}
\begin{aligned}
\call_{\rm linear}=&\frac14M^2_{\rm P}\,\tilde K_{AB}\left(\d l^A\wedge *\d l^B+H^A\wedge *H^B\right)-M^2_{\rm P}\,\tilde K^{IJ}_{i\bar\jmath}\d z_I^i\wedge *\d\bar z_J^{\bar\jmath}\\
&-\frac\ii2 M^2_{\rm P}\left(\tilde K^{ I}_{Ai} \d z^i_I-\tilde K^{I}_{A\bar\imath} \d \bar z^{\bar\imath}_I\right) \wedge H^A\,.
\end{aligned}
\ee
Here the kinetic matrices are obtained by taking double derivatives of the dual potential
\be
\tilde K=K+4\,l^A\Re\rho_A \,,
\ee
with respect to $l^A$, $z^i_I$ and $\bar z^{\bar\jmath}_J$, hence considering $\Re\rho_A$ as function of these fields -- for instance, $\tilde K_{AB}\equiv \frac{\del^2 K}{\del l^A\del l^B}$.

In our case,  the Lagrangian (\ref{dualkinetic}) becomes
\be\label{dualkinetic1}
\begin{aligned}
\call_{\rm linear}=&-4{\rm v}_0a\, M^2_{\rm P}\,\calg_{AB}\left(\d l^A\wedge *\d l^B+H^A\wedge *H^B\right)-\frac{1}{2{\rm v_0} a}M^2_{\rm P}\sum_I g_{i\bar\jmath}(z_I,\bar z_I)\d z_I^i\wedge *\d \bar z_I^{\bar\jmath}\\
&-\ii M^2_{\rm P}\left(\cala^{I}_{Ai} \d z^i_I-\bar\cala^{I}_{A\bar\imath}\d \bar z^{\bar\imath}_I\right) \wedge H^A\,,
\end{aligned}
\ee
where
\be\label{inverseG}
\begin{aligned}
\calg_{AB}&=-M_{{\rm w}AB}+\frac{1}{4{\rm v}_0 a\,}v^Cv^DM_{{\rm w}AC}M_{{\rm w}BD}\\
&=\int_X e^{-4A}\omega_A\wedge *\omega_B
\end{aligned}
\ee
is the inverse of (\ref{calg}).\footnote{In order to prove the second identity first decompose $\omega_A$ in primitive and non-primitive components, $\omega_A=\omega_A^{\rm P}+\alpha_A J$, and then use $*\omega_A^{\rm P}=-J\wedge \omega_A$ and $*J=\frac12 J\wedge J$.}

%%%%%%%%%%%%%%%%%%%%%%%%%%%%%%%%%%%%%%%%%%%%%%%%%%%%%%%%%%%%%%%%%%

\subsection{Rigid limit}

We now consider a decompactification of the above general setting such that $M_{\rm P}\rightarrow\infty$. Recalling (\ref{MP}), we see that the decompactification limit can be obtained by sending ${\rm v}_0\rightarrow\infty$, keeping $a$ and $\Phi$ fixed. 

From (\ref{linearl}) it is clear that the parametrisation of the linear multiplets breaks down in this limit. Hence, it is convenient to rescale them as follows 
\be
l^A\rightarrow -\frac{1}{4{\rm v}_0}l^A\,,\quad~~~~~~~~~~~ H^A\rightarrow -\frac{1}{4{\rm v}_0}H^A\,,
\ee
so that we have the  new  identification
\be
l^A=\frac{v^A}{ a}\,.
\ee
In terms of such rescaled fields the Lagrangian (\ref{dualkinetic1}) becomes
\be\label{dualkinetic2}
\begin{aligned}
\call_{\rm linear}=&-\pi a^2|\Phi|^2\, \calg_{AB}\left(\d l^A\wedge *\d l^B+H^A\wedge *H^B\right)-2\pi|\Phi|^2\sum_I g_{i\bar\jmath}(z_I,\bar z_I)\d z_I^i\wedge *\d \bar z_I^{\bar\jmath}\\
&+\ii \pi a |\Phi|^2\left(\cala^{I}_{Ai} \d z^i_I-\bar\cala^{I}_{A\bar\imath}\d \bar z^{\bar\imath}_I\right) \wedge H^A\,.
\end{aligned}
\ee

On the other hand, after the decompactification, the universal modulus $a$ as well as $\Phi$ become non-dynamical constant parameters. Hence we can actually substitute $(l^A,H^A)$ by new liner multiplets $(v^A,\calh^A)$, with $v^A=a l^A$ and $\calh^A= a H^A$,   and set $\Phi=1$, so that the effective theory becomes
\be\label{dualkinetic3}
\begin{aligned}
\call_{\rm linear}=&2\pi\Big[-\frac12 \calg_{AB}\left(\d v^A\wedge *\d v^B+\calh^A\wedge *\calh^B\right)
-\sum_I g_{i\bar\jmath}(z_I,\bar z_I)\d z_I^i\wedge *\d \bar z_I^{\bar\jmath}\\
&\quad~~~~~~~ +\frac\ii2\left(\cala^{I}_{Ai} \d z^i_I-\bar\cala^{I}_{A\bar\imath}\d \bar z^{\bar\imath}_I\right) \wedge \calh^A\Big]\,.
\end{aligned}
\ee

We can now take the decompactification/$M_{\rm P}\rightarrow \infty$ limit by sending ${\rm v}_0\rightarrow\infty$. Furthermore, we can also take the limit  $a\rightarrow 0$,
which is relevant for the near-horizon geometries considered in the present paper. 
It is clear that generically, in such  limits,   only a subset of  linear multiplets $(v^a,\calh^a)$ remain dynamical and do not decouple. These are selected by the condition that their kinetic terms do not diverge and  remain finite, that is: 
\be\label{normcond}
\calg_{ab}\equiv \int_X e^{-4A}\omega_a\wedge *\omega_b <\infty\,.
\ee
We refer to the harmonic forms $\omega_a$ satisfying (\ref{normcond}) as $L^{\rm w}_2$-normalisable.

Hence, the rigid low-energy effective theory is given by the restriction of (\ref{dualkinetic3}) to the  $L^{\rm w}_2$-normalisable linear multiplets $(v^a,\calh^a)$.  One can then dualise the result back to a rigid supersymmetric Lagrangian using chiral fields $\rho_a$. In fact, one can obtain the dual Lagrangian directly from (\ref{kinetic}), by keeping just the chiral fields $\rho_a$ corresponding to the $L^{\rm w}_2$-normalisable harmonic 2-forms $\omega_a$.  By using (\ref{MP}) and choosing $\Phi=1$ as above, we  obtain
\be\label{dualkinetic4}
\begin{aligned}
\call_{\rm chiral}=-\pi\, \calg^{ab} \nabla \rho_a\wedge *\nabla \bar\rho_b-2\pi\sum_I g_{i\bar\jmath}(z_I,\bar z_I)\d z_I^i\wedge *\d \bar z_I^{\bar\jmath}\,,
\end{aligned}
\ee
where  $\calg^{ab}$  is the inverse of (\ref{normcond}).

\subsection{Inclusion of $B_2$ and $C_2$ axions}

 $C_2$ and $B_2$ moduli can be included along the same lines.  We first need to identify a set of $L_2$-normalisable harmonic forms $\hat\omega_\alpha$, such that 
\be\label{bnorm}
\int_X\hat\omega_\alpha\wedge *\hat\omega_\beta<\infty\,.
\ee
Let us assume that $e^{-4A}$ is at most asymptotically constant as one approaches the boundary of the non-compact $X$.
Then $L^{\rm w}_2$- and $L_2$-normalisable harmonic forms coincide if $e^{-4A}$ is asymptotically constant,
while they can differ when $e^{-4A}$ is asymptotically vanishing, as in the holographic backgrounds considered in this paper. 
In these backgrounds, the   $L_2$-normalisability condition is stronger and the $L_2$-normalisable harmonic forms $\hat\omega_\alpha$ form a subset of the $L^{\rm w}_2$-normalisable harmonic forms $\omega_a$.  Hence, as in section \ref{sec:moduli}, we can split $\omega_a$ in two sets $(\hat\omega_\alpha,\tilde\omega_\sigma)$, where $\tilde\omega_\sigma$ are not $L_2$-normalisable, and expand 
\be
C_2-\tau B_2=\ell^2_{\rm s}(\beta^\alpha\hat\omega_\alpha+\lambda^\sigma\tilde\omega_\sigma)\,.
\ee
The coefficients $\beta^\alpha$ are dynamical moduli entering the four-dimensional effective theory, while $\lambda^\sigma$ are fixed non-dynamical parameters.   
By applying the above rigid limit to  the theory which includes such moduli \cite{Martucci:2014ska} one arrives at the (rigid) effective Lagrangian (\ref{rigidkineticaxion0}).

Notice that the application of the rigid/decompactification limit ${\rm v}_0\rightarrow\infty$ and the near-horizon limit $a\rightarrow 0$ directly on the definition of chiral coordiantes $\rho_a$ (\ref{defchiral}) (completed by the appropriate dependence on the $\beta^\alpha$ moduli \cite{Martucci:2014ska}) and
the K\"ahler potential (\ref{compactK}) is more subtle. For the backgrounds considered in the present paper, it is then easier to directly check that the formulas provided in section \ref{sec:EFT} --  see equations (\ref{rhoa0}) and (\ref{rigidkahler0}) -- give the correct effective Lagrangian.

%%%%%%%%%%%%%%%%%%%%%%%%%%%%%%%%%%%%%%%%%%%%%%%%%%%%%%%%%%%%%%%%%%%%%%%%%%%%%%%%%%%%%%%%%%%%%%%%%%%%
%%%%%%%%%%%%%%%%%%%%%%%%%%%%%%%%%%%%%%%%%%%%%%%%%%%%%%%%%%%%%%%%%%%%%%%%%%%%%%%%%%%%%%%%%%%%%%%%%%%%

%%%%%%%%%%%%%%%%%%%%%%%%%%%%%%%%%%%%%%%%%%%%%%%%%%%%%%%%%%%%%%%%%%

\section{A useful formula}
\label{app:asympt}

Take the cone $C(Y)$ over the Sasaki-Einstein space $Y$. $Y$ can be regarded as a foliation parametrised by the variable $\psi$, whose local transverse space $B$ has a natural K\"ahler structure $j_B$ associated with a transverse metric $\d s^2_B$.  Then the metric on $C(Y)$ can be written as
\be
\d s^2_{C(Y)}=\d r^2+r^2\d s^2_Y\,,
\ee
with
\be
\d s^2_Y=\eta^2+\d s^2_B\,,
\ee
where $\eta$ is the contact form, dual to the Reeb Killing vector. Note that $\d\eta=j_B$ and  $\eta$ can be locally written as
\be
\eta=\d\psi+C\,,
\ee
where $C$ is a locally defined 1-form on $B$, such that $\d C=2j_B$. On $C(Y)$ we can introduce the following vielbein and co-vielbein
\be\label{conevielb}
\begin{aligned}
&E_1=\del_r\,,\quad E_2=\frac1r\del_\psi\,,\quad E_a=\frac{1}{r}\big(e_a-C_a\del_\psi\big)\,, \\
&E^1=\d r\,,\quad E^2=r\eta\,,\quad E^a=r e^a\,,
\end{aligned}
\ee
where $e_a$ ($e^a$), $a=3,\ldots, 6$, is a local (co)vielbein on $B$ and $C_a=\iota_{e_a}C$. Furthermore we can choose a co-vielbein $e^a$ such that we can write
\be\label{Jvielb}
J=\frac12\d(r^2\eta)=r\d r\wedge \eta+r^2 j_B=E^1\wedge E^2+E^3\wedge E^4+E^5\wedge E^6\,.
\ee

Consider now a conical non-compact divisor $D\simeq \mathbb{R}^+\times \Sigma$, with $\Sigma\subset Y$ and conical induced metric $\d s^2_D=\d r^2+r^2\d s^2_\Sigma$. 
We would like to express in a more useful form the quantity
\be
\calj(D,r_{\rm c})=\int_{Y_{\rm c}}\d \text{vol}_YJ\lrcorner \delta^2(D)
\ee
where  $Y_{\rm c}\equiv \{r=r_c\}$ is the transversal five-dimensional slice  isomorphic to $Y$. 
 We can then make the following manipulations
\be
\begin{aligned}
\calj(D,r_{\rm c})=&\frac12\int_{Y_{\rm c}} \eta\wedge j_B\wedge j_B[J\lrcorner \delta^2(D)]\,=\frac1{2r^5_{\rm c}}\int_{Y_{\rm c}} E^2\wedge J\wedge J[J\lrcorner \delta^2(D)]\,\\
=&-\frac1{2r^5_{\rm c}}J^{mn}\int_X\iota_m\iota_n[\delta^1(Y_{\rm c}) \wedge E^2\wedge J\wedge J ]\wedge \delta^2(D)\,.
\end{aligned}
\ee
Now, since $\delta^1(Y_{\rm c})=\delta(r-r_{\rm c})\d r$ and $\delta^2(D)$ has legs along $E^3,\ldots, E^6$, we arrive at
\be
\begin{aligned}
\calj(D,r_{\rm c})=\frac{1}{r^5_{\rm c}}\int_X \delta^1(Y_{\rm c}) \wedge E^2\wedge J\wedge \delta^2(D)=\frac{1}{r^2_{\rm c}}\int_\Sigma \eta\wedge j_B\,,
\end{aligned}
\ee
where  $\Sigma_{\rm c}\equiv D\cap Y_{\rm c}\simeq \Sigma$. We can regard $\Sigma$ as a foliation with transversal holomorphic curve $\calc\subset B$, with metric $\d s^2_\Sigma=\eta^2_\Sigma+\d s^2_\calc$, where $\eta_\Sigma=\d\psi+C|_\calc$. This implies that
\be\label{conediv}
\calj(D,r_{\rm c})=\frac{1}{r^2_{\rm c}}\,\text{vol}(\Sigma)\,.
\ee

%%%%%%%%%%%%%%%%%%%%%%%%%%%%%%%%%%%%%%%%%%%%%%%%%%%%%%%%%%%%%%%%%%%%%%%%%%%%%%%%%%%%%%%%%%%%%%%%%%%%
%%%%%%%%%%%%%%%%%%%%%%%%%%%%%%%%%%%%%%%%%%%%%%%%%%%%%%%%%%%%%%%%%%%%%%%%%%%%%%%%%%%%%%%%%%%%%%%%%%%%

%%%%%%%%%%%%%%%%%%%%%%%%%%%%%%%%%%%%%%%%%%%%%%%%%%%%%%%%%%%%%%%%%%

\section{CS contribution to the E3-brane action}
\label{sec:CSterm}

In this appendix we discuss the CS contribution to the on-shell E3-brane effective action used in section \ref{sec:baryonvev} to compute the vev of baryonic operators.  The CS terms are given by
\be\label{CSaction}
\frac{1}{2\pi}S^{\rm CS}_{\rm E3}=\ell^{-4}_{\rm s}\left(\int_D C_4+\int_DC_2\wedge \calf +\frac{1}{2}\Re\tau\int_D\calf\wedge \calf\right)+\frac1{24}\Re\tau\chi(D)\,,
\ee
where the last term comes from the curvature correction $\ell^{-4}_{\rm s}\int_D C_0\left[\frac{\hat A(TD)}{\hat A(ND)}\right]^{1/2}$ \cite{Green:1996dd,Cheung:1997az,Minasian:1997mm}.

The term $\int_D C_4$ is particularly subtle because of the presence of the D3-branes, which makes $F_5$ non-closed. Hence our strategy will be  to focus on the other terms and to complete the result by  holomorphy. First, $\int_D\calf\wedge \calf$ can be expanded  as in the subsection \ref{sec:DBIcontr}. Furthermore, we can expand  $\int_D C_2\wedge \calf$ in the same way, by using that fact that  we can write   $\int_DC_2\wedge \calf=\int_DC^{\rm h}_2\wedge \calf$, where $C^{\rm h}_2$ is the $L_2$-normalisable harmonic representative of $C_2|_D$ \cite{Martelli:2008cm}.  

By requiring an appropriate pairing with the DBI-terms of subsection \ref{sec:DBIcontr}, it turns out that we must set
\be
\int_D C_4=\ell^4_{\rm s}\,\tilde\varphi +\frac12\int_D B^{\rm h}_2\wedge C^{\rm h}_2
\ee 
where $\tilde\varphi$ naturally pairs with the  $\frac12\int_{D} e^{-4A} J\wedge J$ term in the DBI-action into an $SL(2;\mathbb{Z})$-invariant contribution.   By expanding  $\int_D B^{\rm h}_2\wedge C^{\rm h}_2$ as  $\int_D\calf\wedge \calf$ and $\int_DC^{\rm h}_2\wedge \calf$, 
we arrive at
\be\label{CSexpansion}
\begin{aligned}
\frac{1}{2\pi}S^{\rm CS}_{\rm E3}=&\, \tilde\varphi+\frac{1}{2\Im\tau} \cali^D_{\alpha\beta}\Re\beta^\alpha\Im\beta^\beta+\frac{1}{2\Im\tau}\cali^D_{\alpha\sigma}(\Re\beta^\alpha\Im\lambda^\sigma+\Im\beta^\alpha\Re\lambda^\sigma) \\
&+ \frac{1}{2\Im\tau}\cali^D_{\sigma\rho} \Re\lambda^\sigma\Im\lambda^\rho+\frac{1}{24} \Re\tau\,\chi(D)+I^{\rm CS}_F({\bf f})\,,
\end{aligned}
\ee
where 
\be
I^{\rm CS}_F({\bf f})\equiv  \cali^D_{kl}(\hat N^k{}_\alpha \Re\beta^\alpha+   \tilde N^k{}_\sigma \Re\lambda^\sigma)f^l+\frac{1}{2}\Re\tau \cali^D_{kl} f^k f^l\,.
\ee

The complete E3 effective action is given $S_{\rm E3}=S_{\rm E3}^{\rm DBI}+\ii S^{\rm CS}_{\rm E3}$. 
The requirement  that this combination depends holomorphically on the HEFT chiral fields  singles out the following completion of  (\ref{rhoa0}):
\be
\begin{aligned}
n^a\rho_a=&n^a\Big[\frac12 \sum_I \kappa_a(z_I,\bar z_I;v)+\frac{\ii}{2\Im\tau}\cali_{a\alpha\beta}\beta^\alpha\Im\beta^\beta
\\ &+\frac{\ii}{2\Im\tau}\cali_{a\alpha\sigma}(\beta^\alpha\Im\lambda^\sigma+\lambda^\sigma\Im\beta^\alpha)  \Big]+\frac\ii{2\pi} \sum_I\Im\log\zeta_D(z_I)+\ii\,\tilde\varphi-\ii \varphi\,,
\end{aligned}
\ee
where $\varphi$ is a real constant.  By reabsorbing it in the phase of $\zeta_D(z)$, we can then write
\be
\begin{aligned}
S_{\rm E3}({\bf f})=&\,2\pi n^a\rho_a -\sum_I\log\zeta_D(z_I)+\log r^{\Delta_\calb}_{\rm c}+2\pi I_F({\bf f})+ \frac{\pi \ii}{12}\tau\,\chi(D) +\frac{\pi\ii}{\Im\tau}\cali^D_{\sigma\rho} \lambda^\sigma\Im\lambda^\rho\,,
\end{aligned}
\ee
where $I_F({\bf f})\equiv I^{\rm DBI}_F({\bf f})+\ii I^{\rm CS}_F({\bf f})$, that is
\be
I_F({\bf f})\equiv \ii\cali^D_{lk}(\hat N^k{}_\alpha \beta^\alpha+   \tilde N^k{}_\sigma \lambda^\sigma)f^l+\frac{\ii}{2}\tau \cali^D_{kl} f^k f^l\,. 
\ee

%%%%%%%%%%%%%%%%%%%%%%%%%%%%%%%%%%%%%%%%%%%%%%%%%%%%%%%%%%%%%%%%%%%%%%%%%%%%%%%%%%%%%%%%%%%%%%%%%%%%
%%%%%%%%%%%%%%%%%%%%%%%%%%%%%%%%%%%%%%%%%%%%%%%%%%%%%%%%%%%%%%%%%%%%%%%%%%%%%%%%%%%%%%%%%%%%%%%%%%%%

\end{appendix}

%\bibliographystyle{abe}
%\bibliography{references}{}

\providecommand{\href}[2]{#2}\begingroup\raggedright\endgroup

%%%%%%%%%%%%%%%%%%%%%%%%%%%%%%%%%%%%%%%%%%%%%%%%%

\end{document}